\documentclass[reprint,amsmath,amssymb,amsthm,aps]{revtex4-2}
\usepackage{graphicx}
\usepackage{dcolumn}
\usepackage{bm}
\usepackage{xcolor}
\usepackage{float}

\usepackage{stackengine}
\newcommand\xrowht[2][0]{\addstackgap[.5\dimexpr#2\relax]{\vphantom{#1}}}

\begin{document}
\preprint{}

\title{Model for self-organized Leidenfrost rotating polygons as cnoidal waves}
\author{A. S. Carstea}
\email{acarst@theory.nipne.ro}
\affiliation{Department of Theoretical Physics \\
National Institute of Physics and Nuclear Engineering \\
Bucharest-M\u{a}gurele 077125, Romania}
\author{A. Ludu}
\email{ludua@erau.edu}
\affiliation{Department of Mathematics \\
Embry-Riddle Aeronautical University, Daytona Beach, FL 32114 USA}
\date{\today}

\begin{abstract}
The remarkable appearance of self-organized regular and peaked polygonal rotating patterns in shallow Leidenfrost rings is investigated as a balance between surface tension geometry and nonlinear terms  of Euler equation. Using the Boussinesq shallow convection approximation and a specialized expansion of the Laplace equation solutions, we derive a nonlinear equation that can be integrated in terms of elliptic functions. The model rigorously accounts for surface tension, the contribution of the poloidal rolling vortex, and the interplay between buoyancy-driven and thermocapillary flows. We obtain cnoidal waves solutions describing the dynamics of the inner free surface of the Leidenfrost ring, to predict polygonal patterns in liquid nitrogen. These predictions are compared with experimental observations.Additionally, we introduce a simplified model based on poloidal averaging of the capillary pressure, leading to a Korteweg-de Vries-type equation. This simplified model not only reproduces the cnoidal wave solutions but also predicts new trigonometric solutions, offering insights into the formation of peaked polygons.
\end{abstract}

\maketitle

\section{Introduction}
\label{secintro}

The emergence of regular geometric patterns in compact systems serves as a fascinating example of nonlinear dynamics in fluid mechanics. The formation of rotating polygonal patterns has been studied in various contexts, with theoretical models developed to explain the phenomenon. Experiments and theoretical stability analyses on the formation of polygonal patterns in rotating liquids are presented in \cite{bohr2013,bohr2019,bohr2011,jansson2006}, while the formation of polygonal hydraulic jumps is discussed in \cite{labouse2015,2012bohr}. Experimental studies on the formation of polygonal shapes with $3$ to $5$ edges, or star-shaped patterns, in rotating liquids were conducted to investigate the impact of viscosity on instability development in a shallow layer \cite{abderrah2017}, as well as the effects of sloshing \cite{vatistas90}. 

Among the physical systems exhibiting such features, one interesting example is given by the formation of hollow rotating polygons due to the Leidenfrost effect.  Leidenfrost drops \cite{bia,bohr2019,santos,paul,quere,28,oscillon} or tori \cite{22,28} of various sizes, as well as liquid drops placed on super-hydrophobic surfaces \cite{tam} display self-organized oscillations, star-shaped traveling waves, or other regular patterns due to the constant input of thermal energy, continuous evaporation, and flow beneath the drop. The large variety of patterns formed on these drops and tori results from the competition between surface tension, temperature gradients, gravity, fluid density and velocity  \cite{gelder,yimb,ryu,wei,natu,tam,limbee,chakr,zhang}.
In some of these configurations the axial symmetry of the free surface is broken resulting in  polygonal shapes as a result of the capillary instability of the liquid boundary. 
 
A volatile liquid drop placed on a surface significantly above its boiling point can remain intact for several minutes, owing to the thermally insulating vapor layer that forms underneath the drop \cite{quere,28,labouse2015,chakr,ma}. In this levitated state, often referred to as the Leidenfrost regime \cite{16}, the supporting vapor layer sustained by the evaporation of the liquid, and the liquid is free to undergo frictionless motion due to the absence of liquid–solid contact.

When larger Leidenfrost drops are placed freely on a planar substrate, the underlying vapors accumulates and disturb the shape violently \cite{bia}, or the drop is randomly propelled and escapes observation \cite{yama,linke}.  To address this problem and levitate larger quantities of liquid than the traditional Leidenfrost effect in \cite{28} the authors introduce a substrate that has a circular trough, which allows to stabilize a  liquid torus and obtain nonlinear periodic waves propagating along this levitated liquid ring.

In \cite{amauche2013,hamid}, the authors present nonlinear theoretical models and experiments for shallow layers of Leidenfrost liquid confined in rotating cylinders. They observe the formation of polygonal patterns, traveling cnoidal wave patterns, and hollow-core vortices. Soliton-like profiles of the free surface of the Leidenfrost liquid were also observed in configurations with a circular trough substrate \cite{28,oscillon}. In all these studies the interplay between nonlinearity in the fluid dynamical equations and the geometry appears to be crucial for the appearance of self-organized structures and patterns in such compact systems \cite{book,first}. 

Leidenfrost drops and tori represent a paradigmatic system whose dynamics are of significant interest, not only from a theoretical perspective, but also for industrial applications and natural processes, including mixing \cite{13}, vortices rings \cite{27,qvor}, pharmaceutical industry \cite{31,36}, universal scaling laws for free-surface flows \cite{13} and more. Such studies are also crucial to numerous drop-based technologies, where the Leidenfrost effect can prevent efficient heat transfer, such as spray combustion, or the spray cooling of high performance  microelectronics and optoelectronics devices, or radiological devices \cite{microel1,31,36,naff}. Recent attention has been given to fascinating new topics, such as the self-propulsion of drops on ratchet surfaces and hydrodynamic drag reduction \cite{linke}. Since its discovery by Leidenfrost in 1756 \cite{16} this effect influenced the development of almost all traditional areas of science, and also shaped modern fields of research like microgravity biology, or electromagnetic confinement of plasma. Other applications exploiting the Leidenfrost phenomenon are printing, jet impingement, and vitrification of biofluids by levitation, etc. (for a recent review see \cite{recent}).

  The spontaneous formation of self-sustained, star-shaped oscillations in Leidenfrost drops levitate on a cushion of evaporated vapor over a heated surface is a well-established phenomenon that has been extensively studied and documented in the scientific literature \cite{quere,bia,natu,santos}.  Similar to the Leidenfrost stars, studies of spontaneous oscillations of drops levitated above an air cushion also reveal a breaking of the axial symmetry and the appearance of star drops shapes  \cite{bouw}.   Polygonal instabilities have been observed and reported in a variety of hydrodynamic systems across a wide range of scales including axial symmetric hydraulic jumps \cite{2012bohr} and rotated fluids \cite{bohr2019,bohr2011,bohr2013,jansson2006,abderrah2017}. It was determined that the formation of a roller vortex is a prerequisite for the formation of the polygonal patterns, but  its role was not taken into account in the pressure distribution in these models. In \cite{labouse2015} the authors investigated polygonal formation and instability across a wide range of Reynolds and Weber numbers, and highlighted the critical role of the surface tension and the roller vortex in the polygonal   linear   instability. Nevertheless, in many cases, the mechanism of instability remains poorly understood.   

Recently,  the emergence of solitons was investigated on such circular geometries, including Leidenfrost tori \cite{28}, the free surface of confined rotating flows \cite{hamid,vatistas90,amauche2013}, vortices in super-fluid helium \cite{qvor} and Bose-Einstein condensates \cite{bec}. A possible soliton type of excitation was observed rotating around a Leidenfrost liquid nitrogen drop above a viscous fluid surface \cite{oscillon}. The formation of solitons traveling around closed (spherical) surfaces was predicted first time in 1992, as large amplitude collective excitations of the surface of heavy nuclei in exotic radioactivity processes  \cite{first}, and later on, as shape solitons on the surface of liquid drops \cite{alprl}. Formation of periodic nonlinear waves as cnoidal solutions for the Korteweg-de Vries (KdV) models on closed and bounded liquid systems were also predicted and observed as   rotating hollow polygons in \cite{labouse2015,rag,amauche2013,28,bohr2019,jansson2006,bohr2011}  . Large scale cnoidal waves were also observed as vortex waves, \cite{vortexfriedland}, rotating polygons in the hurricane eye wall, \cite{hurric}, Saturn's North Pole hexagon, \cite{saturn}, and r-modes in neutron stars \cite{neutronstar}.

In this paper we present a nonlinear model for the formation of hollow rotating polygonal patterns in liquid nitrogen Leidenfrost rings and compare the model predictions with our previous experiments \cite{rag}. Our model extends the earlier model proposed in \cite{28}, where the authors obtained a KdV type of equation starting from the Bernoulli equation combined with the capillary pressure contribution.   The nonlinear envelope solution in \cite{28} is represented by a periodization (by construction) of the shape, combined with a spatial cutoff at an azimuthal distance corresponding to a one-soliton envelope of sech$^2$-type.     This theoretical result is important and offers a   reasonable    match with the experiment. Moreover, the envelope of such rotating sech$^2$-envelope decays rapidly and insures the necessary pseudo-periodicity condition.   Similar theoretical approach was used in modeling heavy nuclei or 2-dimensional liquid drops \cite{alprl,first}. The   geometry of the substrate in \cite{28} confines the liquid tori in a circular channel, while in our case the Leidenfrost ring is supported from outside by the vertical wall of the ring, and it develops the polygonal patterns freely, at the inner surface.    The authors in \cite{28} also highlight the formation of a rolling poloidal vortex, an effect that was later investigated in other studies and appears to be crucial for the formation and stability of such rotating polygons.   

In reference \cite{vatistas90} it was shown  the occurrence of rotating regular patterns in a cylindrical container with  water in forced rotation. The resulting patterns are star-shaped with rounded corners. In \cite{hamid} the liquid ring is also in forced rotation regime and the experiments put into evidence solitons orbiting around the circle. In \cite{amauche2013} the authors provide a more comprehensive nonlinear theoretical model for self-organized patterns in free rotation and they obtain analytically $sn$-type of nonlinear waves. However, the cnoidal solutions do not contain the two necessary free parameters of present in the definition of these  Jacobi elliptic functions (period parameter, and independent coordinate scaling parameter) and the predicted shapes are similar to star-shaped patterns or concave polygons very rounded corners.

For liquid nitrogen Leidenfrost rings on flat substrate no complete theory exists so far to predict the spontaneous formation of sharp corner polygons in rotation, modeled as cnoidal waves solutions of a fully nonlinear equation.  In that, our model represents  an  effort    in this direction, using a more precise expression for the surface tension, and including the rolling vortex contribution in a specific way. Not all parameters in our model can be obtained from first principles, and we therefore use four free parameters which are derived from the experiments. We model the dynamics of the inner free surface of the Leidenfrost ring using the Euler equation for an incompressible liquid in the Boussinesq shallow convection approximation. The fluid velocity is decomposed in potential and rotational components and the corresponding Helmholtz-Hodge boundary conditions are applied. The potential component of the velocity field is solved using the Laplace equation for a power series expanded in the radial coordinate. Because the liquid layer is considered very shallow, we are able to approximate the vorticity with a gradient, and to include it in an integrable equation. In this way, the resulting evolution equation for the free inner surface of the liquid ring can be combine with the boundary conditions in one nonlinear differential equation for the contour curve $\gamma(\theta,t)$. Because the goal of the paper is to predict  the formation of rotating polygonal patterns inside the liquid ring, we solve this equation in the co-rotating frame, in one single variable $\xi=\theta-\Omega t$. Vorticity is implemented in the model in an integral form, in terms of the vortex Reynolds number, which later on is considered a free matching parameter. The surface tension term is calculated in detail using the mean curvature for the parametrization of the inner surface $\Sigma$, and the very long formulas are expanded in power series of small parameters which depend on the aspect ration of the liquid ring. The final form of the evolution equation contains 12 terms, with nonlinearity considered up to   order $\mathcal{O}(\gamma^3)$ in the function describing the inner contour of the liquid  . This equation is solved approximate analytic, and the emerging cnoidal waves solutions (elliptic Jacobi functions) are analyzed. The cnoidal solutions are matched with the experimental data for polygonal patterns using four free parameters, and a special procedure for treating the variation of the parameters with respect to the poloidal angle $\phi$ of the vertical profile of the surface. An interposed section is devoted for the validation of this model by averaging the surface tension term, and solving a simplified version of the evolution equation. 

The layout of the paper is as follows. In section  \ref{secmodel0}, we provide a detailed description of our Leidenfrost ring system, including its geometry and the characteristic fluid velocities. We identify the temperature regime and examine the potential for turbulent flow. We introduce the thermal properties of liquid nitrogen and investigate the onset of Rayleigh-B\'{e}nard-Marangoni instability. Additionally, we discuss the balance between buoyancy-driven and thermocapillary flows. Finally, we evaluate and analyze several dimensionless numbers (Reynolds, Bond, Prandtl, Galileo, Weber, Biot, Marangoni, Rayleigh, and Grashof) relevant to the fluid dynamics of our system. In section \ref{secmodelNL}, we present the geometric parametrization of the liquid's free surfaces. The core of the paper is laid out in section \ref{seceqnl}, where we introduce the dynamical model, analyze the potential flow and vorticity, apply the boundary conditions, and derive the evolution equation  Eq. (\ref{eq2}). In Section \ref{rolvort}, we outline our approach for the vorticity component, describe the structure of the rolling vortex, and examine the term responsible for capillary pressure. The detailed calculation of the surface tension term can be found in Appendix I. In section \ref{qualanal}, we incorporate the vorticity and surface tension terms into the evolution equation Eq. (\ref{eq5}) and provide a qualitative analysis of this nonlinear differential equation. The family of cnoidal wave solutions to the evolution equation is introduced in section \ref{semodel1}, with detailed calculations provided in Appendix II. In section \ref{secmodel2}, we present a simplified version of the evolution equation, where an averaged surface tension term is used over the poloidal angle. As a form of model validation, we show that this simplified equation can be mapped to a Korteweg-de Vries-type integrable equation, which yields the same cnoidal wave solutions, albeit in a different parametrization. Finally, in section \ref{secexp}, we present the experimental setup and the procedure for comparing experimental data with model predictions. Further details of this procedure are provided in Appendix III.

\section{Qualitative description of the Leidenfrost ring}
\label{secmodel0}

We study the spontaneous formation of rotating polygonal patterns at the interior of Leidenfrost rings. The occurrence of such patterns constitutes a most intriguing phenomena observed at the frontier between vortex dynamics and free-surface flows. Even if the focus of this paper is a theoretical nonlinear for these polygons, in order to introduce working hypotheses, we briefly describe below the experimental setting. More details on the experiments are presented in section   \ref{secexp}. 

\begin{figure}[H]
\centering
\includegraphics[scale=0.4]{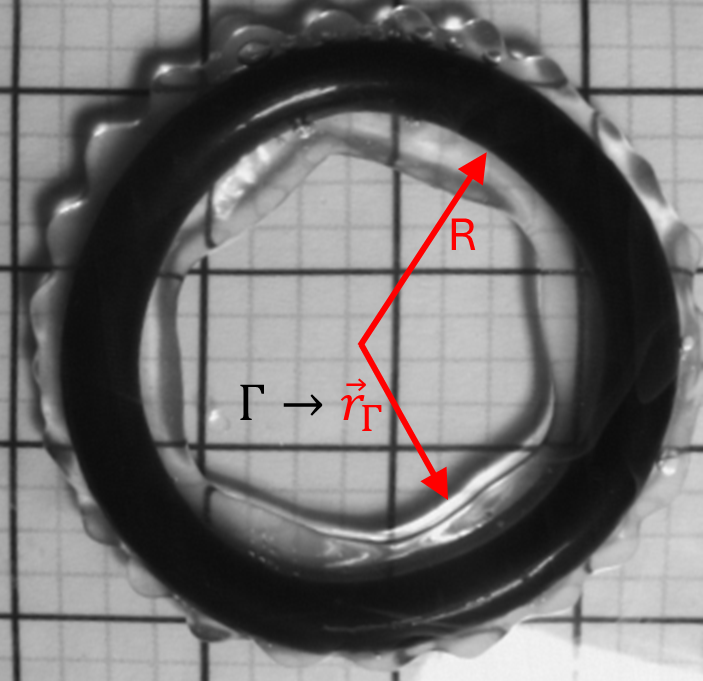} 	
\caption{Typical liquid nitrogen Leidenfrost ring formation inside a rigid cylinder. The liquid layer maintained outside the cylinder creates a regime which favors formation of rotating hollow polygons inside. $R$ is the inner cylinder radius, and $
	\vec{r}_{\Gamma}(\theta,t)$ is the parametrization of the $\Gamma$ directrix describing the inner liquid boundary.}
	\label{fig0}
\end{figure}

The experiment is realized  by  gently  placing liquid nitrogen on a flat horizontal substrate  inside and outside a shallow cylindrical wall  (a ring of inner radius $R=1.5-3.0$ cm)  Figs. \ref{fig0}-\ref{fig1} and \ref{figs6543}-\ref{figs6543b}.   The liquid outside the ring is physically separated from the liquid inside, during the observation of rotating patterns.  The purpose of adding liquid nitrogen outside the ring is to maintain the temperature of the solid ring as close as possible to the boiling point of liquid nitrogen. This thermal regulation is crucial for ensuring consistent experimental conditions.

The bottom of the substrate is maintained at room temperature, which it is above the Leidenfrost temperature $T_{L}\simeq 125.15$ K for liquid nitrogen \cite{quere,bia,labouse2015,28}. The ring is sealed to the substrate and it is in contact with liquid nitrogen at its inside and outside lateral surfaces, so during the experiment the ring  temperature drops close to   $T_b =77$ $K\ll T_L$   Fig. \ref{fig1}, and consequently the liquid nitrogen  adheres to the cylinder walls Fig. \ref{fig0}. The liquid inside the ring levitates over the substrate because of the evaporation-driven lubrication pressure which develops under the liquid \cite{chakr}. Due to Rayleigh-Taylor instability bubbles rise from the vapor layer beneath the liquid \cite{quere,bia,ma} for such a large puddle. To suppress this effect the authors in \cite{quere,ma} use curved substrate surfaces. In our system we eliminate this effect by consecutively applying a number of cycles of pouring and complete evaporation until the ring temperature is brought close to $T_b$. In the last cycle an empty region is formed at the center, and the liquid takes the shape of a ring, with its upper and underneath surfaces being in contact with the vapors, a Leidenfrost state \cite{16,quere}. In this state, the inner boundary of the Leidenfrost liquid which surrounding the empty region, begins to oscillate, exhibit rotational traveling waves and   finally generates almost rigid shape rotating polygonal patterns  Fig. \ref{fig0}, 4 movies (Hexagon1, Hexagon2, Square1, Square2)   and 7 figures (squares: N4.A, N4.B; hexagons: N6.A, N6.B, N6.C; heptagon: N7; and octagon N8)   as Supplemental Materials \cite{supp}. During this process  the upper liquid surface remains almost flat and flush with the solid ring, and the liquid remains in contact with the inner vertical ring wall Figs. \ref{fig0}-\ref{fig1}. The inner surface circular waves continue to propagate until the Leidenfrost ring  evaporates almost completely (details provided  in section   \ref{secexp}.   Similar cases of axial symmetry breaking with formation of nonlinear star-shaped drops are discussed in \cite{bouw} and explained in terms of Rayleigh modes.  

In our configurations, the liquid nitrogen layer becomes unstable because of a vertical temperature gradient created by  higher temperature at the bottom  (warm, Fig. \ref{fig1}) and the heat losses at the inner and upper surfaces due to evaporation. Several studies  \cite{yimb,ma,chakr} show temperature variations within Leidenfrost droplets of a few Kelvin,  while drops with a few millimeters height have almost constant temperature \cite{yimb}. We estimate for our system  ($h=0.3-0.5$ cm)   a vertical change in liquid temperature around   $\delta T\sim 5$ K,   which is sufficient to induce buoyancy-driven and surface-tension-driven convection flows. Such gradients, especially in the neighborhood of the boundaries,  initiate the formation of vertical rolling vortices with the upwards and downward draft concentrated very close to the liquid boundaries.

The unstable modes formed in the ring generate patterns of rotation and oscillations in the inner surface segment $\vec{r}_{\Sigma}$ Fig. \ref{fig1}, and a rolling vortex motion in the vertical plane. The typical frequencies we  observed for such modes are in the range $\Omega = 4.5-11.7$ s$^{-1}$, similar to the modes frequencies for Leidenfrost drops of comparable sizes reported in \cite{yimb}. Such types of symmetry breaking have been extensively investigated in \cite{yama,yimb}, both experimentally and numerically, and the results imply that the rolling motion is likely to be induced by the thermobuoyant (Rayleigh–B\'{e}nard) effect rather than the thermocapillary (Marangoni) effect. 

While the origin of such symmetry breaking is still under discussion, the relevance of rolling motion to the polygonal pattern formation is feasible,   and the coupling between  poloidal vorticity and a free surface it seems to be a generic feature \cite{labouse2015}.   Similar  toroidal vortex roller structure was discussed  in \cite{2012bohr} and \cite{28}. Also, in experiments on zig-zag translation of Leidenfrost droplets \cite{yimb} it was confirmed that such type of rolling motion is crucial to the global translation of droplets. Similar results of  toroidal steady flow were obtained in \cite{yimb}. In \cite{chakr,ma} it is shown how spontaneous symmetry-breaking modes parametrically driven by the pressure variations in the vapor layer can create rotating and oscillating star-shaped patterns in Leidenfrost drops.  Evaluation of the Froude number   $Fr=0.5$    supports the case of a subcritical flow  similar to the downstream flow for similar polygonal patterns in the hydraulic jumps case  \cite{2012bohr}.

\section{Model parametrization}
\label{secmodelNL}

\begin{figure}[H]
\centering
\includegraphics[scale=0.25]{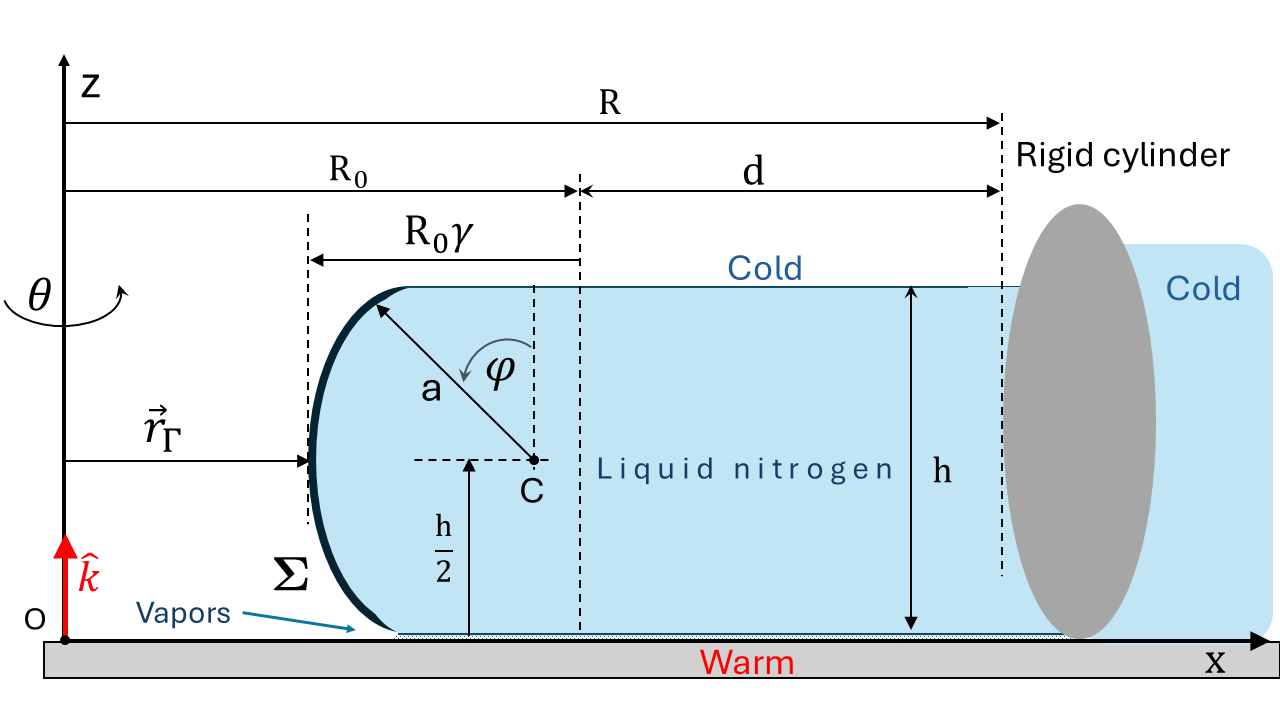} 	
\caption{A normal cross section in the Leidenfrost ring of height $h$. $R_0$ is the  radius corresponding to the static equilibrium ($\gamma=0$, $\Gamma$ circular) and $R$ is the radius of the external solid boundary. The inner boundary of the liquid is the normal channel surface $\Sigma$  approximated  by a circular arc of radius $a$ (thick black curve) and center $C$ which is not necessary placed on the radius $R_0$. $\Sigma$ is parameterized by the contour $\vec{r}_{\Gamma}(\theta,\varphi,t)$.  Liquid surface deformations at the proximity of the wall and bottom are not shown here, and the figure is not drawn to scale. In this example $\gamma<0$ }
\label{fig1}
\end{figure}

We employ cylindrical coordinates. The boundaries of the Leidenfrost ring consist in the vertical region in the proximity of the rigid wall $r=R$, the two approximately flat horizontal surfaces on top and bottom $z=h$, and $z=0$, respectively, and the inner surface $\Sigma$ surrounding the hollow region, Fig. \ref{fig1}. The $\Sigma$ surface has the most important role in the dynamics of the ring-shaped Leidenfrost liquid. Similar Leidenfrost configurations were obtained when the liquid is confined in a circular trough and are called levitated liquid tori \cite{28}.
We build the geometry of our model starting with a  horizontal simple loop $\Gamma$,  called directrix or contour, lying at $z=h/2$ \cite{x3}. Based on this curve we construct  a normal channel surface $\Sigma$ (also called ringed or pipe surface   \cite{ringed,tubu}) generated by parallel sweeping a circular arc of radius $a$ with its middle point on the directrix. Each such circular arc is contained  in the normal plane of $\Gamma$ and has its convexity towards the origin of coordinates. In this way we model the inner capillary surface of the Leidenfrost ring.  A normal cross-section is presented in Fig. \ref{fig1} where the black thick circular arc represents $\Sigma$. We parameterize the time-dependent directrix $\Gamma$   parameterized by the polar, plane curve $\gamma(\theta,t)$   versus the azimuth angle $\theta \in [0,2 \pi)$ around the $z-$axis
\begin{equation}\label{ega}
\vec{r}_{\Gamma}=R_0 [1+\gamma(\theta,t)]   \vec{e}_r +\frac{h}{2} \vec{e}_{z},  
\end{equation}
  where $\vec{e}_r=(\cos \theta, \sin \theta, 0)$ and  $\vec{e}_{z}=(0,0,1)$  are the radial and  $z-$axis unit vectors, respectively, (see Fig. \ref{fig1}).  
Here the parameter $R_0<R$ is the static equilibrium radius of the liquid in the absence of the deformation, given by the equilibrium liquid volume would be in absence of oscillation Vol$\simeq\pi (R^2 -R_0^2) h$, where we neglected here the contribution to volume of the circular vertical profile of radius $a$. We assume that $R-R_0=d$ and $|\gamma(\theta,t)|$ are small quantities compared to $R$. Based on Eq. (\ref{ega}) we parameterize $\Sigma$ by the azimuth and poloidal angles $(\theta,\varphi)$ in the form \cite{x3,tubu,ringed,book}
\begin{equation}\label{A1e1}
\vec{r}_{\Sigma}=\vec{r}_{\Gamma}(\theta,t)+a [\vec{n}_{\Gamma} (\sin \varphi-1)+(1+\cos\varphi) \vec{e}_{z} ],
\end{equation}
where $\vec{n}_{\Gamma}$ is the unit normal  to the curve $\Gamma$ parameterized  as in Eq. (\ref{ega}). In general, for an arbitrary circular arc we have $\varphi \in [\varphi_{min},\pi-\varphi_{min}]$ with $\varphi_{min}=\arccos h/(2a)$ or $h=2a\cos \varphi_{min}$,   (see Fig. \ref{fig1}).   In the limiting case when the circular arc is a   semicircle,   we have $h=2a$, $\varphi_{min}=0$. An example of such a surface is presented in Fig. \ref{fig2} for a hexagonal directrix $\Gamma$. Knowing the true vertical profile of the inner surface $\Sigma$ is a complex problem because the local geometry of the contact line depends on the variation of curvature because of liquid oscillations ,  on the temperature, and on the shear stress created by the vapor flow. There are precise calculation in literature where profiles are given by integral equations \cite{shapy}, yet for the shallow rings model we consider the approximation of the profile with circular arcs to serve the result.  
The present shape parametrization for shallow Leidenfrost rings ($a\ll R_0$) is similar to parametrization for Leidenfrost drops used before in literature \cite{2012bohr,28,chakr}.

\section{Nonlinear model for the Leidenfrost ring}
\label{seceqnl}

We can consider  liquid nitrogen as inviscid in a good approximation  because both its kinematic and dynamic viscosity, and the Prandtl number Pr$\simeq 0.2$ are one order of magnitude smaller than the values for water \cite{liqnitr}.   Although the Reynolds number for the system under consideration does not exceed $10^4$ (in our cases $1800<$Re$<3700$), it remains relatively large in the context of the flow dynamics. Moreover, the Eckert number Ec$=V^2/(c_p \delta T)\sim 6 \cdot 10^{-6}\ll 1$ with $c_P=2 \cdot 10^3 J/$(Kg K) for liquid nitrogen and $dT\simeq 5-10$ K, $V\sim 0.2$ m$\cdot$ s$^{-1}$ is very small, indicating that viscous dissipation effects are negligible. Additionally, this study focuses on the bulk motion of the fluid (such as oscillations and spinning), rather than on fine-scale boundary layer phenomena.   The Richardson number associated to shallow Leidenfrost rings $Ri=Gr/Re^2\simeq 0.14$ is relatively small, so the natural convection is not dominant process for this system. Not only the thermal difference in inertia of the liquid  can be negligible, but even the buoyancy is not  strong in the bulk since the critical liquid height   $L_{crt}=0.1 C_p (\delta T)/(\alpha_v g T_{b})$, (according to Eq. 41 in D. D. Gray et al \cite{ri})   above which we need to include buoyancy  in the evolution equation according the criteria \cite{ri}, is  at least one order of magnitude larger that the Leidenfrost ring height   $h=0.3-0.5$ cm.    It is thus reasonable to use the Boussinesq approximation for shallow convection (Oberbeck-Boussinesq) and to consider the liquid nitrogen in our system as inviscid and incompressible. Consequently,  the dynamics can be described by  the incompressible Euler equation without gravitational potential energy 
\begin{equation}\label{emat1}
\frac{\partial \vec{V}}{\partial t}+(\vec{V} \cdot \nabla) \vec{V}=-\frac{\nabla P}{\rho},
\end{equation}
for the velocity field $\vec{V}$, pressure $P$ and density $\rho$. From the incompressibility condition $\nabla \cdot \vec{V}=0$ Eq. (\ref{emat1}) can be written
\begin{equation}\label{emat2}
\frac{\partial \vec{V}}{\partial t}+\nabla \biggl( \frac{ P}{\rho}+\frac{|\vec{V}|^2}{2}\biggr)=\vec{V} \times \vec{\omega},
\end{equation}
where $\vec{\omega}=\nabla \times \vec{V}$ denotes the vorticity \cite{v2}. Assuming the velocity field of the liquid is sufficiently smooth and its domain of definition is bounded with regular boundary \cite{bookchorin}, the Helmholtz-Hodge Decomposition theorem provides for the velocity a unique decomposition in a gradient field with prescribed normal on the boundary, and a divergence-free vector field parallel to the boundary \cite{helmh2b,h4} 
\begin{equation}\label{emat3}
\vec{V}=\nabla \Phi+ \nabla \times \vec{A}, \ \  \vec{N}_{\Sigma} \cdot (\nabla \times \vec{A})=0, 
\end{equation}
where $\Phi$ is the velocity potential obeying $\triangle \Phi=0$, $\vec{Q}=\nabla \times \vec{A}$ is a solenoidal field and $\vec{N}_{\Sigma}$ is the normal to the liquid  boundary.  Another way to apply the decomposition theorem would be to ask the gradient field to be normal on the boundary \cite{h2,helmh2b}. This option is not convenient for our system because we are looking for horizontal rotation modes of  the free inner liquid surface, and consequently this boundary has a non-zero azimuthal (tangent) component of the velocity. 

We plug Eq. (\ref{emat3}) in Eq. (\ref{emat2}) and obtain
$$
\nabla \biggl( \frac{P}{\rho}+\frac{\partial \Phi}{\partial t}  +  \frac{|\nabla \Phi|^2}{2}  +  \nabla \Phi \cdot \vec{Q}+|\vec{Q}|^2 \biggr) 
$$
\begin{equation}\label{emat4}	
+\frac{\partial \vec{Q}}{\partial t}-\nabla \Phi \times \vec{\omega}  -  \vec{Q} \times \vec{\omega}=0.
\end{equation}
We can use a gauge transformation $\vec{A}\rightarrow \vec{A}+\nabla \Psi$ and choose the scalar $\Psi$ such that $\nabla \cdot (\vec{A}+\nabla \Psi)=0$  while the solenoidal velocity $\vec{Q}$ remains the same. In this case $\vec{\omega}=-\triangle \vec{A}$ and $\vec{A}$ can be obtained as the Green integral solution for the vector Poisson's equation. 

Because the Leidenfrost rings analyzed here are very shallow   ($h=0.3-0.5$ cm $\ll R$),   the rolling vortices tend to form in the vertical plane, and then continue to flow along the radial direction, hypothesis which is corroborated with our experiments. In our observations of the flow  we note  upwards flow  at the free inner boundary $\Sigma$ and downwards flow closed to the rigid boundary of the wall, as expected from the values of the Biot and Grashof numbers, as discussed  in section II.  From these observations we infer that the dominant component of the vorticity is along the azimuthal direction $\vec{e}_{\theta}$ and $\vec{\omega}\simeq \omega \vec{e}_{\theta}$. For the same reason
we can neglect in the following the variation of the velocity in the vertical direction.

It results that the only non-zero component of $\vec{A}$  is the azimuthal one, similar to $\vec{\omega}$, and since we neglect the $z$ dependence of the velocity, $\vec{Q}$ is oriented along the $z$ direction and $\nabla \Phi \cdot \vec{Q}=0$ in Eq. (\ref{emat4}). Another consequence is that $\vec{Q}\times \vec{\omega}$ has only a radial component, while $\vec{Q}$ has only the vertical component $Q\vec{e}_z$.  
We introduce an auxiliary potential function $\Psi=z \omega \Phi_r$. Because $|z|\le h\ll R$ is small we can approximate $\nabla \Psi\simeq \nabla \Phi \times \vec{\omega}\simeq (0,0,\omega \Phi_r)$ in cylindrical coordinates. With these observations  Eq. (\ref{emat4}) becomes
\begin{equation}\label{e1}
\nabla \biggl( \frac{P}{\rho}+\frac{\partial \Phi}{\partial t}+\frac{|\nabla \Phi|^2}{2}-\Psi\biggr)=\vec{Q} \times \vec{\omega}-\frac{\partial \vec{Q}}{\partial t},	
\end{equation}
where we neglected the higher order nonlinear term $|Q|^2$ from Eq. (\ref{emat4}).
The velocity is expressed now in terms of its radial, azimuthal and vertical components given by $\vec{V}=(\Phi_{r},\Phi_{\theta}/r,Q)$ under the hypotheses mentioned above.

Because the boundary conditions for the rolling vortex flow $\vec{N}_{\Sigma}\cdot \vec{Q}=0$ are already considered through the Helmholtz-Hodge Decomposition, we consider the boundary condition for the potential component of the velocity on the inner cylindrical rigid  surface in the form 
\begin{equation}\label{e01}
\Phi_r=0 \ \ \ \hbox{on} \ \ \ r=R_0+d. 	
\end{equation}
  If the boundary cylinder has straight vertical inner wall, this is an exact relation. If the cylinder is an elliptic cross section torus,  we can still neglect its vertical curvature $h\ll 2 R$ because it would introduce only a correction in the second order in the further calculations.  
The nonlinear kinematic boundary condition for the potential velocity on the free surface $\Sigma$ is 
\begin{equation}\label{e02}
\dot r \equiv \Phi_{r}=R_0 \biggl(\gamma_t+\frac{\gamma_{\theta}\Phi_{\theta}}{r^2} \biggr)  \ \ \ \hbox{on} \ \ \ \Sigma. 
\end{equation}
A linearization of the free surface boundary condition  $\Phi_r=R_0\gamma_t$ would allow only radial oscillations, which is not the object of this model.  

Following the perturbation technique described in \cite{whit} we introduce a power series for the velocity potential
\begin{equation}\label{e2}
\Phi=\sum_{n=0}^{\infty}\left(\frac{r-R_0}{R_0}\right)^n f_n(\theta,t),
\end{equation}
with the radius of convergence controlled by the $r/R_0-1=\gamma$ small amplitude variable \cite{book}. 

We introduce the series Eq. (\ref{e2}) in the expression of the Laplacian $\triangle \Phi$ in cylindrical coordinates, and evaluate the result on the free surface $\Sigma$,  Eq. (\ref{ega}). In this form the Laplacian can be expanded in a similar power series in terms of $\gamma$ on the surface $\Sigma$
\begin{equation}\label{e3}
\begin{split}
& \triangle \Phi |_{\Sigma}  =
\sum_{n\ge 0} \biggl[  \frac{n (n-1) \gamma^{n-2} f_n}{R_{0}^2} \\ 
& + (-\gamma)^{n}  \sum_{k\ge 0}^{n} \frac{(-1)^k k f_k  
+   (1+n-k) f_{k,\theta\theta}}{R_{0}^2} 
\biggr]
\end{split}
\end{equation}
where we used the expansions
\begin{equation}\label{e4}
\frac{1}{1+\gamma}=\sum_{n\ge 0}(-\gamma)^n, \ \ \frac{1}{(1+\gamma)^2}=\sum_{n\ge 0}(1+n)(-\gamma)^n.
\end{equation}
Since in this paper we focus on understanding the dynamics of contours $\Gamma$ having the shapes of  rotating regular polygons, we can estimate the validity of the series approximations in Eq. (\ref{e2}-\ref{e4}). In the polygonal cases, $r$  ranges between the inner radius $r_{in}$ and the circumradius $r_{out}$ of any $n-$regular polygon, so $r_{out}-r_{in}$ gives the upper bound for $|\gamma|$, and $r_{in}<R_0 <r_{out}$. If we truncate these series only to order $\mathcal{O}(\gamma^2)$ the relative truncation error  lies between $0.05\%$ for $n=8$ and $4\%$ for $n=4$. Moreover, even a linear approximation in $\gamma$ does not exceed $12\%$ relative truncation error for $n=4$, with $2\%$ relative truncation error for a hexagon, $n=6$. 
 
If the order of truncation in all the series exceeds $2$, Eq. (\ref{e3}) can always be written in the first two orders in $\gamma$ in the form
\begin{equation}\label{e5}
\begin{split}
\triangle \Phi |_{\hbox{on} \ \Sigma} & =\frac{f_1+2 f_2+f_{0,\theta\theta}}{R_{0}^2} \\
& + \gamma \frac{2 f_2 -f_1 +6 f_3 -2 f_{0,\theta \theta}+f_{1,\theta \theta}}{R_{0}^{2}}+\mathcal{O}(\gamma^2)=0.
\end{split}
\end{equation} 
From the outer surface boundary condition and Eq. (\ref{e01}) and Eq. (\ref{e2}) we have
\begin{equation}\label{e6}
\sum_{n\ge 1}\frac{n d^{n-1}}{R_{0}^{n}} f_n=0, \ \hbox{or} \ f_2=-\frac{R_0 f_1}{2d}+\mathcal{O}\biggl(\frac{d^2}{R_{0}^2}\biggr).
\end{equation} 
Using Eqs. (\ref{e2}, \ref{e4}) in the inner surface kinematic boundary condition  Eq. (\ref{e02})  we have
\begin{equation}\label{e7}
R_{0}^{2} \gamma_t = \sum_{n\ge 0}\biggl( n f_n-\gamma \gamma_{\theta} \sum_{j=0}^{n} (-1)^{n-j}(1+n-j)f_{j,\theta}\biggr),
\end{equation} 
and showing the first orders
\begin{equation}\label{e7}
f_1-R_{0}^{2} \gamma_t=\gamma_{\theta} f_{0,\theta}-2 \gamma f_2+\gamma \gamma_{\theta} (f_{1,\theta}-2f_{0,\theta})+\mathcal{O}(\gamma^3).
\end{equation}
We process this equation using the perturbation technique \cite{whit}, and notice that when the nonlinear terms are neglected in the original Eq. (\ref{e02}),  we may write the linearization of Eq. (\ref{e7}) in the form \cite{lamb}
\begin{equation}\label{e8}
f_1=R_{0}^{2} \gamma_t.
\end{equation}
Retaining the terms of second order in $\gamma$  in Eq. (\ref{e6}), and using Eq. (\ref{e8}) and the expansions from Eq. (\ref{e4}) we obtain after some algebraic manipulation 
\begin{equation}\label{e9}
\frac{d}{R_0} f_{0,\theta} \gamma_{\theta}+R_{0}^{2} \gamma \gamma_t=\mathcal{O}(\gamma^3).
\end{equation}
Introducing Eqs. (\ref{e8},\ref{e9}) in the first term of  Laplace equation Eq. (\ref{e5}) we obtain in the leading order in $\gamma$ and $d/R_0$ 
\begin{equation}\label{e10}
f_2  =-\frac{R_{0}^{2} \gamma_t}{2}+\frac{R_{0}^3}{2d}\biggl(\frac{\gamma \gamma_t}{\gamma_{\theta}}\biggr)_{\theta}.
\end{equation}
We obtained that, in the working approximation, the leading terms in the velocity potential Eq. (\ref{e2}) can be expressed only function of the contour function $\gamma(\theta,t)$. Namely, the term $f_0$ in Eq. (\ref{e9}), $f_1$ in Eq. (\ref{e8}) and $f_2$ in Eq. (\ref{e10}). These terms  are also crucial for the leading order of the azimuthal and radial velocities.   A similar approach  for the potential flow is used in modeling the formation of star-shaped drops levitated by an airflow \cite{bouw}.  

In the following, we differentiate the azimuthal component of Eq. (\ref{e1}) with respect to $\theta$ and evaluate it on the inner surface $\Sigma$ 
\begin{equation}\label{euler}
\frac{\partial}{\partial \theta} \biggl( \Phi_{t}+\frac{\Phi_{r}^{2}}{2}+\frac{\Phi_{\theta}^{2}}{2r^2} +\frac{P}{\rho} -z\omega \Phi_r \biggr)_{\hbox{on} \ \Sigma}=0.
\end{equation}
Using Eqs. (\ref{e8}, \ref{e9}), we obtain in the leading orders in $\gamma$ and $d/R_0$ the following strongly nonlinear equation:
\begin{equation}\label{eq2}
\begin{split}
&- \left(\frac{R_0^3\gamma\gamma_t}{d\gamma_{\theta}} \right)_{t}+R_{0}^{2} \gamma_t \gamma_{t \theta} \\
&-\biggl[\frac{R_{0}^{4}}{2 d^2} \biggl( \frac{\gamma \gamma_t}{\gamma_{\theta}} \biggr)^2+h\omega R_0 \gamma_t \biggr]_{\theta}+\biggl( \frac{P_{\theta}}{\rho} \biggr)_{\hbox{on} \ \Sigma} =0. \\
\end{split}
\end{equation}
In the following we analyze the contribution of the vorticity and surface tension terms of this equation.

\section{Contribution of rolling vortex flow and surface tension}
\label{rolvort}

In order to evaluate the vorticity $\omega$ associate with the Rayleigh-B\'{e}nard-Marangoni rolling vortex flow we use Kelvin's Theorem, and we choose a vertical cross section along the radial direction, similar to the blue shape in Fig. \ref{fig1}, to the left of the rigid boundary ring. We assume that the rolling flow  is due to the congruence of thermocapillary and buoyancy forces, and it is mainly localized at the boundary of the liquid, and tangent to the surface, as discussed above. As mentioned in \cite{2012bohr} these rolling vortices are a strong idealization of the complex flow occurring in these systems. In order to  include them in the model we rely on the observation that the vorticity is mainly azimuthal. Consequently, we can integrate $\omega$ on this vertical cross section and calculate it as the circulation   $\Gamma_{K}$   by the Stokes Theorem for the Kelvin contour velocity
\begin{equation}\label{vorty}
\Gamma_{K}=\iint \omega dS=   \oint   \vec{V} \cdot d\vec{r}\simeq 2 V_z h+2 V_r (R-r_{\Gamma}).
\end{equation}
For our system the gradient of density is vertical and the gradient of pressure (dominated by the surface tension with Bo$\sim 2-5$) is azimuthal, the baroclinic vector $\nabla P \times \nabla \rho$ lies in the radial cross section and following Kelvin's Theorem (precisely Ertel's Theorem) the circulation conserves in this surfaces. The surface  flow of this roller structure generates a toroidal vortex similar to the one described in \cite{2012bohr} for polygonal hydraulic jumps where Bo$=2.3$. According to this reference the instability which favors the formation of polygons occurs for sufficiently small Bond numbers. A similar result on the formation of coherent vortices of uniform vorticity with toroidal roller structure  was obtained for geophysical flows on the surface of a sphere \cite{potvor}. 

Following our experimental data we noticed that the vertical component of the velocity at the free boundary $\Sigma$ is slightly larger than the radial velocity on the horizontal boundaries. To include the rigorous thermocapillary and buoyancy dynamics in our present model would require more detailed knowledge about the vertical profile of $\Sigma$ and its temperature distribution.   This inclusion will be addressed in a forthcoming publication, where it will be examined within the framework of a more refined model.   Since the goal of the present model
is only to predict the formation of the hollow rotating polygons, we introduce the vortex Reynolds  number   $Re_{\Gamma}=\Gamma_K/\nu$   \cite{vortre} as our scaling parameter, which is unknown, but using the fit with the experiments we can estimate its value. Therefore we consider that the tangent velocity $Q_{\Sigma}$ along the vertical part of the Kelvin contour bounding the radial cross section can be expressed as $Q_{\Sigma}\simeq Re_{\Gamma}\Phi_{r,surface}\simeq Re_{\Gamma} R_0 \gamma_t$. 

The average value of vorticity in the radial cross sections can be obtain from the conservation of the quantity in Eq. (\ref{vorty})
\begin{equation}\label{v1}
\omega=2\biggl( \frac{Re_{\Gamma} R_0}{R-R_0}+\frac{R_0}{h}+\frac{Re_{\Gamma} R_0^2 \gamma}{(R-R_0)^2}\biggr)\gamma_t
\end{equation}

The geometric nonlinearity occurs in Eq. (\ref{eq2}) through the surface tension term $P_{\theta}$. The capillary pressure generated by surface tension is provided by the Young-Laplace equation \cite{batch}
\begin{equation}\label{e11}
	P-P_0=-2\sigma H_{\Sigma},
\end{equation}
where $H_{\Sigma}$ is the mean curvature \cite{x3,book} of the $\Sigma$ inner liquid surface, $\sigma$ is the coefficient of surface tension, and $P_0$ is the vapor pressure next to the liquid surface. 
\begin{figure}[h]
	\centering
	\includegraphics[scale=0.39]{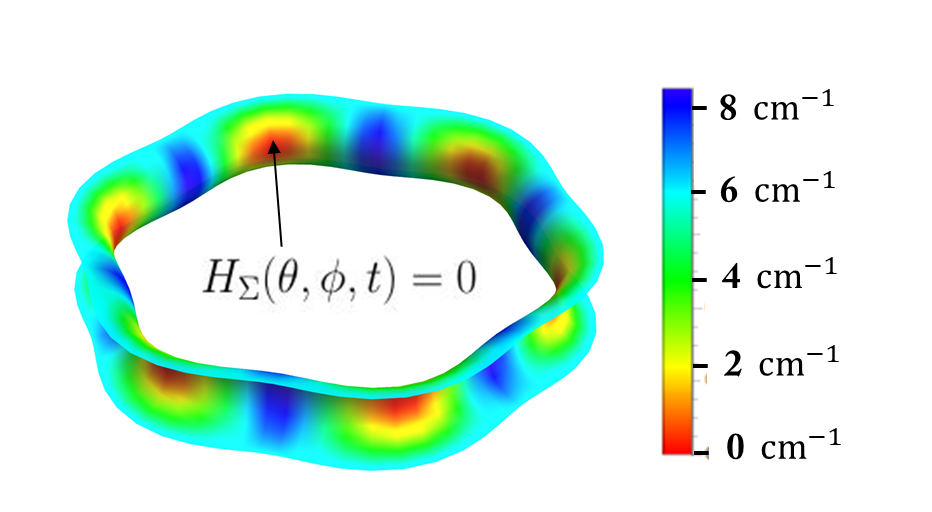} 
	\caption{Example of an inner liquid boundary $\Sigma$ generated by a hexagonal shape $\gamma$.  The mean curvature $H_{\Sigma}$,  Eq. (\ref{eqp}) is represented in colored density plot, with legend. The corners of the contour $\Gamma$ have large positive curvature $\kappa_{\Gamma}$ and can be compensate the large half-cylinder negative curvature $-1/a$ of the tubular surface, resulting in $H_{\Sigma}\simeq 0$ around these critical points (red regions).}
	\label{fig2}
\end{figure}
The expression of the capillary pressure from Eq. (\ref{e11}) is too long to be presented here, but we describe the  procedure to calculate it in Appendix 1. Eqs. (\ref{eqp}-\ref{A1e4}) in Appendix 1 provide the expression of the mean curvature in a truncated form up to order $4$ in the smallness parameters $\epsilon=a/R_0$ and $\delta=(a/R_0)\sin \varphi$. This truncation  Eq. (\ref{eqp}) has the form
\begin{equation}\label{eqhcu}
	H_{\Sigma}  = \sum_{k=0}^{3}A_k \gamma^k+B_1 \gamma_{\theta}^{2}
	+B_2 \gamma_{\theta \theta} +C_1 \gamma \gamma_{\theta \theta}+C_2 \gamma \gamma_{\theta}^{2},	
\end{equation}
where the coefficients $A,B,C$ are given in Eqs. (\ref{A1e4}) in terms of $\epsilon, \delta$ and $a$.

\section{The nonlinear evolution equation}
\label{qualanal}

Our goal is to identify the existence of uniformly rotating rigid patterns. 
The evaluation of the Womersley number $Wo\simeq 10^3$ (sometimes denoted in literature    $\alpha=$Wo$=R_0 (\Omega/\nu)^{1/2}$ and defined as the dimensionless expression of the pulsatile flow frequency $\Omega$ in relation to viscous effects   ) for our system indicates that the flow is dominated by oscillatory inertial forces, and the velocity profile is flat.  For Womersley parameters above 10, the unsteady inertial forces dominate, and the flow is essentially one of piston-like motion with a flat velocity profile \cite{womers}. Accordingly, we feel justified to introduce in this model the co-moving orbital coordinate (sometimes called traveling-wave reduction) $\xi=\theta-\Omega t$ and Eqs. (\ref{eq2},\ref{eqhcu}-\ref{v1}) combine into our final evolution equation
\begin{equation}\label{eq5}
\begin{split}
& -\biggl( \frac{\Omega^2  R_0^3}{d}+\frac{2 \sigma A_1}{\rho}\biggr)\gamma'
 +\biggl(\frac{\Omega^2R_0^4}{d^2}-\frac{4\sigma A_2}{\rho}\biggr)\gamma\gamma' \\
& -\frac{6\sigma A_3}{\rho}\gamma^2\gamma' 
 -\biggl[3 \Omega^2 R_{0}^{2}+\frac{4\Omega^2 R_{0}^{2} Re_{\Gamma} h}{d}+\frac{4\sigma B_1}{\rho} \biggr] \gamma'\gamma''	\\
& -\frac{2\sigma B_2}{\rho}\gamma'''-\frac{2\sigma C_1}{\rho}(\gamma \gamma'')' \\
&-\biggl[ \frac{2 Re_{\Gamma} h\Omega^2 R_{0}^{3}}{d^2}+\frac{2\sigma C_2}{\rho}\biggr] (\gamma \gamma'^{2})'=0,
\end{split}
\end{equation}
where $\gamma_{\xi}=\gamma'$   and the coefficients $A_i, B_i, C_i$ are calculated in Appendix 1, Eqs. (\ref{A1e4}).   We note that for the thin layer approximation  ($\epsilon, \delta \ll 1$) the contribution of the radial velocity in the evolution equation is not quite critical. This component of velocity is represented by the coefficient $\Omega^2 R_{0}^{4}/d^2$ of the $\gamma \gamma'$ term in  Eq. (\ref{eq5}), but this term  is  present in this evolution equation in any case through the other coefficients.

In the order $\mathcal{O}(\gamma^4)$ the evolution Eq. (\ref{eq5}) has a family of periodic waves approximate analytic solutions in the form of the Jacobi elliptic function cn, known as cnoidal waves \cite{lamb,draz,first,book}
\begin{equation}\label{solcno}
\gamma(\xi)=s_2+(s_3-s_2){\rm cn^2}[p(\xi-s_3)|m],
\end{equation}
where the family parameters depend on the physical parameters of the system $We, R_0, d, h, \rho,  \sigma$ , and also on four free parameters $Re_{\Gamma}$, $a_0$ integration constant from Eq. (\ref{eq6}), and $\epsilon, \delta$ from Eq. (\ref{A1e3p})) in Appendix section \ref{secapp2}.

Eq. (\ref{eq5}) can be written in dimensionless form using the azimuthal Weber number   $\mathcal{W}e=\rho v^2 L/\sigma$   as the ratio between rotational kinetic energy and surface tension energy, where we plug $v=\Omega R_0$ and $L=2 \pi R_0$, that is $\mathcal{W}e=2 \pi \rho \Omega^2 R_0^3 /\sigma$. We also introduce the dimensionless aspect ratio $=h/d$ of the liquid ring, the smallness parameters $\epsilon=a/R_0, \delta=\epsilon \sin \phi$ in Eq. (\ref{A1e3p}), and the filling factor $d/R_0$ (Fig. \ref{fig1}).   Using these notations, we present an abbreviated form for Eq. (\ref{eq5}), where we expressed  the surface tension coefficients $A_i,B_i,C_i$  only up to order $\delta^2$ for simplicity 
\begin{equation}\label{eq5xx}
\begin{split}
& \biggl( -\frac{\epsilon \mathcal{W}e }{2 \pi}\frac{R_0}{d}+\delta \biggr)\gamma'
+\biggl[  \frac{\epsilon \mathcal{W}e }{2 \pi}\biggl(\frac{R_0}{d}\biggr)^2-2\delta  \biggr]\gamma\gamma' \\
& +3 \delta \gamma^2\gamma' 
-\biggl[ \frac{3\epsilon \mathcal{W}e }{2 \pi}+\frac{8 \epsilon^2 Re_{\Gamma} \mathcal{W}e}{2\pi}\frac{R_0}{d}+\delta  \biggr] \gamma'\gamma''	\\
& +\delta \gamma'''-2\delta (\gamma \gamma'')' \\
&+\biggl[ -\frac{2\epsilon^2 Re_{\Gamma} \mathcal{W}e }{\pi}\biggl(\frac{R_0}{d}\biggr)^2+\frac{3 \delta}{2} \biggr] (\gamma \gamma'^{2})'+\mathcal{O}(\delta^2)=0.
\end{split}
\end{equation}
In all subsequent calculations we use the full expressions Eq. (\ref{eq5}) with complete expressions for where the surface tension coefficients $A_i,B_i,C_i$ from Eqs. (\ref{eqp}, \ref{A1e4}). This strongly nonlinear equation depends on four free parameters $Re_{\Gamma}, a_0, \epsilon, \delta$, the last two defined in Eq. (\ref{A1e3p}). These parameters are related to the rolling vortex induced by the thermocapillary/buoyancy effect and to the vertical profile of the inner boundary $\Sigma$. All other parameters occurring in Eq. (\ref{eq5}): $\mathcal{W}e$ (or $\Omega$), $R_0, d, h$ can be determined  from the experimental setting.  
\begin{figure*}[h!tbp]
	\centering
	\includegraphics[scale=.45]{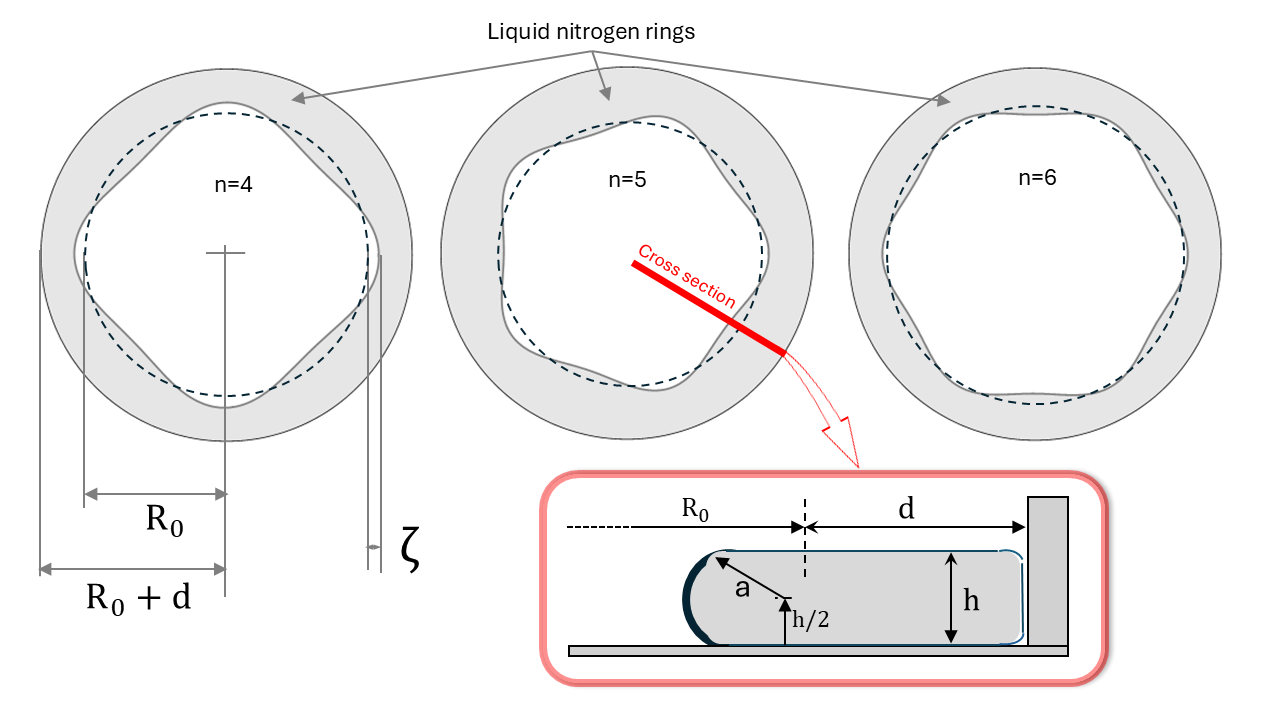} 	
	\caption{Examples of cnoidal waves solutions Eq. (\ref{el}) representing hollow polygons in the co-rotating frame, with $4,5$ and $6$ edges. All geometric parameters occurring in Eq. (\ref{eq5}) and the amplitude of the contour deformation $\zeta$ are shown.  The polygons with $n=4,5,6$ edges are also presented in Figs. \ref{figs6543} left, right and Fig. \ref{figs6543b} left, respectively.  }
	\label{figtheo}
\end{figure*}

The three traditional terms of the classical KdV equation \cite{lamb,perturb} are grouped in Eq. (\ref{eq5xx}) in a reduced expression
$$
\biggl( -\frac{\epsilon \mathcal{W}e }{2 \pi}\frac{R_0}{d}+\delta \biggr)\gamma'
+\biggl[  \frac{\epsilon \mathcal{W}e }{2 \pi}\biggl(\frac{R_0}{d}\biggr)^2-2\delta  \biggr]\gamma\gamma' +\delta \gamma'''\simeq 0.
$$ 
Even if in the following we use the full form of the evolution Eq. (\ref{eq5}), we note that this reduced form can still provide rotating cnoidal wave solutions (solitary waves) in $\mathcal{O}(\epsilon,\delta)$. This reduced form is useful for a quantitative  analysis of the structure of its three coefficients. In order to provide cnoidal waves solutions they should have the appropriate signs,  namely 
$$
 -\frac{\epsilon \mathcal{W}e }{2 \pi}\frac{R_0}{d}+\delta <0, \ 
  \frac{\epsilon \mathcal{W}e }{2 \pi}\biggl(\frac{R_0}{d}\biggr)^2-2\delta  >0.
$$ 
These conditions request the value for the azimuthal Weber number $\mathcal{W}e>2 \pi d/R_0\simeq \pi$. We will show in section \ref{secexp} that in all our experiments $d/R_0\simeq 0.5$ resulting in values $\mathcal{W}e>3.5$, which completely fulfill this condition. Moreover, since this azimuthal Weber number is related to the angular velocity, this constraint requests $\Omega\ge 4.5$ rad/s which condition is fulfilled in all observed rotating hollow polygonal shapes. 

It is useful to make a comparison between our model Eqs. (\ref{eq5},\ref{eq5xx}) and the classical KdV equation for shallow water gravity waves with surface tension, over flat and horizontal bottom \cite{lamb,alprl}. In the classical KdV case, traveling soliton solutions, or cnoidal waves are possible if the coefficients of two out of the three dominant terms have the same sign, namely in our notation the nonlinear term $\gamma \gamma'$ and the linear dispersion term $\gamma'''$. Usually the time derivative term (in our case $\gamma'$) must have the opposite sign. In the flat KdV case this is guaranteed if the coefficient of surface  tension $\sigma$ doesn't exceed a certain critical value, or if the Bond number is $Bo<1/3$ (see for example Eq. (11.20) in section 11.1 in \cite{book}). This situation is possible because of an appropriate balance between gravity and surface tension interactions. In our circular model there is no gravity involved in the evolution Eq. (\ref{eq5}) since the nonlinear deformations happen in the horizontal plane. The linear dispersion term $\gamma'''$  has always positive sign. In this case the dominant nonlinear term $\gamma \gamma'$ in $\mathcal{O}(\gamma^2)$ has positive coefficient only because of the contribution of the factor $\mathcal{W}e R_0^2$ which means the classical balance between gravity and surface tension is substituted here by the balance between the curvature $R_0$ and the rotation $\mathcal{W}e$ against the surface tension contribution. And similarly to the flat bottom KdV case, in our model the coefficient of the $\gamma \gamma'$ surface tension term should be less than a critical value, namely $We$ should be larger than a certain value $\gtrsim \pi$. This observation may bring some understanding about the observed nonlinear rotation waves, as sharp polygonal shapes at the inside.  

In the following we present the cnoidal waves solutions of Eq. (\ref{eq5}) for two typical geometric configurations: when $h\ll a$ (section III) and when $h\simeq a$ (section IV).  Such analysis helps to understand the interplay between the geometric nonlinearity and the physical nonlinearity given by convective terms of the  momentum balance which root from Eq. (\ref{e1}).

\section{Cnoidal waves solutions. Case 1}
\label{semodel1}

In this section ,   we consider the configuration of a shallow ring  $h,a \ll R_0$ where the $\Sigma$ vertical profile of the inner boundary is rather flat and it has a small curvature $h=2 a \cos \varphi_{min} \ll a$, Fig. \ref{fig1}. In this case the formation of  nonlinear cnoidal waves solutions  Eq. (\ref{eq5}) as rotating polygons is not critically controlled by dependence on the poloidal angle $\varphi$ in the $\delta$ parameter since  the range of  $\varphi$ along $\Sigma$ is narrow, and we are able to use Eqs. (\ref{eqp}-\ref{A1e4}) for mean curvature and capillary pressure considering $\delta$ a constant parameter. We are interested in the evolution of the inner boundary contour $r_{\Gamma}=R_0(1+\gamma(\theta,t))$ provided by the cnoidal waves   \cite{lamb}      solutions of Eq. (\ref{eq5}) in the form of Eq. (\ref{e20}) from Appendix 2 
\begin{equation}\label{el}
\gamma(\theta,t)=s_2+(s_3-s_2){\rm cn^2}[p(\theta-\Omega t-s_3)|m].
\end{equation}
  Similar cnoidal waves solutions were predicted or observed on systems like liquid drops, surface of heavy nuclei, vortex systems, 2-dimensional electron droplets in magnetic field, etc. \cite{first,alprl,vortexfriedland,dors,book}.  
In the following we use the procedure described in detail in Appendix II. We can skip the constant term $s_3$ inside the argument of the Jacobi elliptic cosine function because its relevance is related to an arbitrary choice of the initial moment of the rotation. The roots $s_1,s_2,s_3$ are the roots of the polynomial Eq. (\ref{eqpolyn}) and  the scaling factors $p,m$  are given in   Eq. (\ref{eqparame}) and depend on  these roots and one of the coefficients of the polynomial. 

Since $r_{\Gamma}$ is a bounded and periodic function we define its limits with the help of an amplitude $\zeta$,  $0\le \zeta<d$ representing the maximum radial deformation of $\gamma$ with respect to $R_0$, or $R_0-\zeta\le r_{\Gamma} \le R_0+\zeta$ and $|\gamma|\le \zeta/R_0$,  see for example Fig. \ref{figtheo}. Using these limits and Eq. (\ref{el}) we obtain the relationship between shape and the polynomial roots 
\begin{equation}\label{201}
s_1=-\frac{\zeta}{R_0}+\frac{2\zeta}{R_0}\biggl(1-\frac{1}{m}\biggr), \ s_{2,3}=\mp\frac{\zeta}{R_0},  
\end{equation}
\begin{equation}\label{202}
\gamma(\theta,t)=-\frac{\zeta}{R_0}+\frac{\zeta}{R_0}{\rm cn^2}[p(\theta-\Omega t)|m].
\end{equation}
 
Since  the cnoidal waves Eq. (\ref{202}), or Eqs. (\ref{el}-\ref{201}), are solutions of the nonlinear dynamical Eq. (\ref{eq5}), the shape and kinematics parameters $R_0,\zeta,p, m, \Omega$ of these solutions are in direct correspondence with the coefficients of Eq. (\ref{eq5}). 
It follows that, for a given choice of $R_0$ and $\Omega$, one can fit a specific rotating polygonal contour $\Gamma$ by appropriately selecting the physical coefficients in the main evolution Eq. (\ref{eq5}) . Thus, the resulting polygonal shapes are determined by measurable physical parameters of the system, as well as by the four model parameters  $Re_{\Gamma}, a_0$ and $\epsilon, \delta$. More details on these calculations are given in Appendix 2, specifically Eqs. (\ref{203}, \ref{222}).

In Fig. \ref{figtheo} we present examples of cnoidal waves solutions  Eq. (\ref{el})  where the parameters $\zeta, p,m$ are chosen to display three typical polygonal shapes.

In the hypothetical limit $m\rightarrow 1$, the cnoidal solution Eq. (\ref{202}) approaches an aperiodic one-soliton profile of of sech$^2$, which is confirmed by the models in \cite{28,amauche2013}. An increase in the value of $m$ for given fixed geometry $R, d, R_0, \epsilon,\delta, h$ is possible by in increase in the angular rotation $\Omega$, that is the azimuthal Weber number. Faster polygons tend reach their quasi-equilibrium at smaller number of edges, as it is also noticed in Table I.

A theoretical estimation of the rotation velocity $\Omega$ for the polygons can be obtained from the nonlinear dispersion relation for the wave dynamical Eq. (\ref{eq5}) for the cnoidal wave solutions Eq. (\ref{el}, \ref{202}). The dispersion relation is obtained using the averaging procedure described in  \cite{whit,book}, and in more detail in \cite{kevr}, and we obtain the dependence of the angular velocity $\Omega(n; h, R, a, \delta, d, \zeta)$ on the geometry of the polygon and the other parameters of the configuration. The dependence is presented in  Eq. (\ref{204x}),  Appendix 2.

More calculation details for the matching procedure provided by Eqs. (\ref{201}-\ref{203}) and the properties of the cnoidal solution are presented in Appendix 3.

\section{Cnoidal waves solutions. Case 2: averaged pressure}
\label{secmodel2}

The second case we analyze considers the vertical cross section in the liquid ring to be rather rounded than flat, when $h\simeq a\ll R$. In this case it is acceptable to retain the leading linear terms $\mathcal{O}(\gamma)$ in the expression of the mean curvature $H$  Eqs. (\ref{eqhcu},\ref{eqp}) 
$$
P=-2\sigma(A_0 + A_1 \gamma+B_2\gamma_{\theta \theta})+\mathcal{O}(\gamma^2).
$$ 
We  average the dependence on the azimuthal variable $\phi$ in these coefficients 
\begin{equation}\label{p1}
<P>=-\frac{2\sigma}{\pi}\int_{0}^{\pi}(A_0+A_1\gamma+B_2\gamma_{\theta \theta})d\phi=P_0+P_1 \gamma+ P_2 \gamma_{\theta \theta}
\end{equation}
and using the expressions of the coefficients from Eqs. (\ref{A1e4}) we obtain
\begin{equation}\label{a2}
\begin{split}
& P_0=\frac{2\sigma}{\pi}\frac{\pi R_0+2 \pi \sqrt{R_0^2-a^2}+2R_0 \arctan\frac{a}{\sqrt{R_0^2-a^2}}}{2 a \sqrt{R_0^2-a^2}},  \\
& P_1=\frac{\sigma R_0 \biggl(-a \pi+2 \sqrt{R_0^2-a^2}+2a\arctan \frac{a}{\sqrt{R_0^2-a^2}} \biggr)}{\pi (R_0^2-a^2)^{\frac{3}{2}}},  \\
& P_2=\frac{\sigma \biggl( -2R_0 +\frac{\pi R_0 a}{\sqrt{R_0^2-a^2}}-\frac{2R_0 a}{\sqrt{R_0^2-a^2}}\arctan \frac{a}{\sqrt{R_0^2-a^2}}\biggr)}{\pi (R_0^2-a^2)}.  \\
\end{split}	
\end{equation}
Introducing these averaged pressure terms in the system   Eq. (\ref{eq2}) and using Eq. (\ref{v1}) in the same way we proceeded in section \ref{rolvort},  we obtain in the co-moving orbital coordinate $\xi$ a new (and simpler) expression for the evolution, similar to  Eq. (\ref{eq5}) previously obtained. Differentiating this new equation with respect to $\xi$ we find the following nonlinear differential equation 
\begin{equation}\label{finn}
-\biggl(\frac{\Omega^2 R_0^3}{d}+\frac{P_1}{\rho}\biggr)\gamma'+\frac{\Omega^2R_0^4}{d^2}\gamma\gamma'+\frac{P_2}{\rho}\gamma'''=0
\end{equation}
This equation is exactly the traveling-wave reduction of the Korteweg de Vries equation. The solution of Eq. (\ref{finn}) is given by the family of periodic cnoidal waves 
\begin{equation}\label{eqfin2}
\begin{split}
& \gamma(\xi)=\frac{4 d^2 P_2 p^2(1-4k-4k^2)-d(\Omega^2 R_0^3\rho+ P_1 d)}{\Omega^2 R_0^4\rho} \\
& +\frac{4 d^2 P_2 kp^2(2 +k )}{\Omega^2 R_0^4\rho}{\rm cn}^2(p\xi|k),	\\
\end{split}
\end{equation}
where $k,p$ are free parameters for the family. This solution Eq. (\ref{eqfin2}) is belongs to the same family of waves as the one obtained in section \ref{semodel1}, Eq. (\ref{solcno}), except it is represented in a different mathematical parametrization. The conclusion is that KdV cnoidal waves can describe the dynamics and shapes of various regimes of curvature and thickness for the Leidenfrost rings, of course in the limit of shallow rings.  For appropriate choice of its parameters, Eq. (\ref{eqfin2}) generates the same patterns as those obtained previously by the cnoidal waves in Eq. (\ref{el}), and is able to fit our experimental observations equally good. 

In the configuration described in this section, if we consider the contribution of the radial flow in the Euler equation (given by $\Phi_r^2/2$ in the leading order) Eq. (\ref{finn}) becomes
\begin{equation}\label{fi3}
\gamma'\biggl[ \frac{\Omega^2 R_0^3}{d}+\frac{ P_1}{\rho}+\frac{\Omega^2R_0^4}{d^2}\gamma+\Omega^2 R_0^2\gamma''\biggr] +\frac{ P_2}{\rho}\gamma'''=0,
\end{equation}
which  is not reducible to an integrable system. However,  Eq. (\ref{fi3})  admits exact solutions in the form of rotating concave polygonal patterns (having peaked corners) 
\begin{equation}\label{peak}
\gamma=\frac{ P_2 R_0^2 -d\Omega^2 R_0^3 \rho - P_1 d^2  }{\Omega^2 R_0^4\rho}-m\left|\sin\left(\frac{R_0(\theta-\Omega t)}{d}\right)\right|,
\end{equation}
where $m$ is a free parameter. These solutions can generate peaked polygons with a number $n=3,\dots,8$  of vertices when the ration $2R_0/d=n$ is integer, like for example the match with our experiment presented in the right frame of Fig. \ref{figs6543b}.

\section{Comparison with experiments}
\label{secexp}

Experiments are performed at room temperature. Liquid nitrogen is slowly poured  inside and outside a shallow cylindrical ring  Figs. \ref{fig0}-\ref{fig1} and \ref{figs6543}- \ref{figs6543b} sealed on a horizontal regular glass substrate of thickness $1$ cm. After a number of cycles of pouring and complete evaporation of liquid nitrogen, the ring temperature is reduced close the boiling point. In the last pouring (usually the third time), an empty region is formed at the center, and the liquid takes a ring shape, flush with the boundary solid ring, with its upper and underneath surfaces being in contact with its vapors, the Leidenfrost state. 

The evaporation characteristic time  is much larger than the characteristic time of the polygonal patterns rotation. The evaporation-driven dynamics can be decoupled from the free surface pattern formation dynamics, so we can consider $R_0,d$ as constant for a number of rotations of each polygonal pattern. The same conclusion about the stability of Leidenfrost inner flow  was obtained in \cite{yimb} supporting the quasi-static stability hypothesis.

The radial and azimuthal velocities of the fluid,  measured using tracers (see movies as Supplemental Materials \cite{supp}), are in the range $V \sim 0.1-0.2$ m$\cdot$s$^{-1}$, typical for internal-flow velocity of Leidenfrost droplets \cite{yama,yimb}. Our boundary cylinders have inner radii $R=1.5-3$ cm and height   $h=0.3-0.5$ cm,   Fig. \ref{fig1}, consequently the Reynolds number is $Re=1,800-3,700$. Inertia overcomes viscosity (liquid nitrogen $\nu= 2 \times 10^{-7}$ m$^2$ s$^{-1}$) and   it   seems that the regime for Leidenfrost rings is favorable for the formation of polygonal patterns in rotation because the critical Reynolds number for the onset of Leidenfrost shape oscillations as described in \cite{sim} is $Re_{c} \simeq 2,200$,  while the onset of turbulent flow occurs at higher $Re_{c} \gtrsim 2,800$ \cite{liuthermo}.   This situation is similar to the analysis performed in \cite{bouw}, where it was demonstrated that the potential component of this flow is relevant, while the temperature effect are secondary to some extent.   It is interesting to compare the liquid nitrogen Leidenfrost rings case with the formation of polygonal patterns under hydraulic jumps instability, for more viscous liquids (ethylene glycol) where $Re=212$   \cite{2012bohr}. 

The dependence on temperature of the surface tension coefficient  for liquid nitrogen is given by \cite{surften,liqnitr,gra} 
$$
\sigma (T)=\sigma_c \biggl( 1-\frac{T}{126.25} \biggr)^{1.247},
$$
where $\sigma_c=29.09 \times 10^{-3}$ Nm$^{-1}$ is the surface tension at the critical temperature $T_c=126.26$ K. In the following we will denote  the derivative of $\sigma(T)$ with respect to $T$ by  $\sigma'(T)$. At $77$ K we have $\sigma=8.96 \times 10^{-3}$ N m$^{-1}$ and it results a Bond number $Bo=7.94$ and capillary length   $l_c=0.1$cm.   Other parameters for liquid nitrogen were obtained from   \cite{surften,liqnitr,gra}: density $\rho=807$ Kg m$^{-3}$, thermal diffusivity $\alpha_{T}=10^{-6}$ m$^2$s$^{-1}$, thermal expansion coefficient $\alpha_v=9 \times 10^{-3}$ K$^{-1}$, expansion rate liquid-gas at normal pressure is $177$, and thermal conductivity $\lambda=0.134$ W mK$^{-1}$. 
It follows a relatively small Prandtl number $Pr=0.202<1$ which implies that thermal diffusivity is dominant  compared to convection \cite{bia,chakr,liqnitr,gra}.

The deformation of the upper surface of the liquid ($z=h$ in Fig. \ref{fig1})  can be neglected on the scale $\mathcal{O}(h)$ since the crispation number  $Cr=\nu \rho \alpha_T /(\sigma R)\simeq 3.3 \times 10^{-5}$ is small, and the Galileo  number is large  $Ga=g h^3/(\nu \alpha_T)  \simeq 2.37 \times 10^5$ \cite{sak} is large.

Typical values obtained for the angular velocity of polygonal patterns range in $\Omega=4.5-11.7$ s$^{-1}$, values which are associated with Weber number $We=\rho h V_{z}^{2}/\sigma =2-4$ for the vertical flow, and $We=\rho R^3 \Omega^2/\sigma=5-10$  for the azimuthal flow. The kinetic energy of the liquid nitrogen is larger than the surface tension energy \cite{yimb} justifying the tendency of formation of fast rotating rigid patterns. Similar results were obtained for hydraulic jumps generated patterns \cite{2012bohr} where the typical angular velocity is in the range $30$ s$^{-1}$ and $We=1.9$. 


The height of the Leidenfrost ring   $h=0.3-0.5$ cm,   and the corresponding temperature difference in the range of $5$ K induce  buoyancy-driven and surface-tension-driven convection flows. The corresponding Biot number Bi$=2.7>1$ also indicates that the thermal gradients  within the liquid are significant close to the boundaries of the liquid. When both effects are coupled the result generates a  Rayleigh-B\'{e}nard-Marangoni convection instability \cite{zhang,liuthermo}. For our setting the Rayleigh number  $Ra=\alpha_v g R^3 \delta T  /(\nu \alpha_T)\simeq 6 \times 10^4$ and the Marangoni number $Ma=-\sigma' h \delta T /(\rho \nu \alpha_T) \simeq 4 \times 10^4$ greatly exceed the expected critical values for the onset for Rayleigh–B\'{e}nard and Marangoni instabilities,  typically $Ra_{c} \simeq 340$ \cite{liuthermo}, or  $\simeq 10^3$ \cite{yimb}  and $Ma_{c} \simeq 10^2$ \cite{yimb},  for similar geometries. In our system the Biot number is large $Bi\simeq 5$ leading to a lower bound for the critical Marangoni number $Ma_{c}\simeq 120$ \cite{liuthermo}. These observations are in agreement with the formation of vertical rolling vortices inside the Leidenfrost ring with  upwards and downward draft concentrated  on the liquid boundaries (see movies and photographs as Supplemental Materials \cite{supp}).

The Rayleigh number also exceeds the threshold of $1,708$ for buoyancy-driven convection \cite{chakr,limbee,gelder}. We also assume for our system that the buoyancy-driven flows induces stronger surface driven flows than surface-tension-driven flows since the  Marangoni/Rayleigh ratio $Ma/Ra \simeq 0.3<1$ is small \cite{gelder}. 

Buoyancy forces actually augment the thermocapillary flow along the vertical segments of the liquid boundary  Fig. \ref{fig1} \cite{limbee}. The effect of the Rayleigh-B\'{e}nard-Marangoni convection instability is to induce flow with fluid pulling up the free surface from the base (warm) towards the top, as noted in other similar systems \cite{chakr}. Even at this high Marangoni number, the flow associated to a Prandtl number of $Pr\simeq 0.2$ in our case, is not turbulent,  leading to $Re\sim 2,500$ \cite{liuthermo}. 

The thermobuoyant (Rayleigh–B\'{e}nard) effect triggers unstable modes of rotating-oscillating patterns with frequencies in the range $4-12$ s$^{-1}$, similar to those in \cite{yimb},  as well as the formation of poloidal rolling vortex tubes with velocity in the vertical plane \cite{yama,yimb}. In our configuration it appears that the rolling motion is likely to be induced by  rather than the thermocapillary (Marangoni) effect. From the values of the Grashof number $Gr\simeq 2.6 \times 10^5$ and the P\'{e}clet number $Pe \simeq 300$ we note that the convective effect are indeed pronounced \cite{liuthermo}, and the flow is within   a transient regime,   without occurrence of flow separation \cite{gra,bouw}. At this high Grashof number the isothermals are parallel to the cold wall, and the downwards return of the rolling vortex is formed farther from the cylinder wall \cite{naff}. 


The convective heat coefficient for horizontal plates is $h_{T}=0.54 \lambda Ra^{1/4}/L$ where the Rayleigh number is $Ra=\rho \beta g L^3 \delta T /(\mu \alpha)$ and  $\beta=0.009$ is the coefficient of thermal expansion of liquid nitrogen at boiling point. With these values we can calculate the Nusselt number $Nu=h_T L/\lambda \simeq 8.41$ whose value supports the hypothesis   of a transitional flow as a mix of laminar flow and localized bursts of turbulence and intermittent instabilities.   The thermal  conductivity of the silicon material for the ring is $\lambda=2.5$ W (mK)$^{-1}$ \cite{rubber} with a convective rate of $1.87$, and for the glass substrate $\lambda=0.58$ W (mK)$^{-1}$ with a convective rate $4.33$ \cite{thermcondglass}. These values for the thermal transfer parameters provide sufficient support for the substrate to realize a Leidenfrost state underneath the liquid ring, but a adhesion contact with the vertical ring wall.   

The motion of the liquid was recorded with AOS high-speed camera set at $1000$fps with view point straight above the ring, at $10$ cm distance. Fluorescent  tracers  (Cospheric\textsuperscript{\tiny\textregistered} $0.1-100\mu$m, $800$ Kg/m$^3$) were added to the liquid nitrogen to visualize the flow \cite{rag,book}. See the 4 movies as Supplemental Materials: Hexagon1 and Hexagon2, Square1 and Square2,   and also 7 photographs of squares, hexagons, a heptagon  and an octagon   \cite{supp}. The high-speed movies were analyzed frame by frame using "PIVlab 3.0 beta" MATLAB open source software. The shape of the ring provides the parameters $R, h$, and for each observed polygonal pattern we measured the angular speed $\Omega$ (which gives us the value for $We$ parameter), and the parameters $R_0, d$. The other unknown quantities $Re_{\Gamma}, a_0, \epsilon$ and $\delta$ are considered free for the match with the theoretical model. 
\begin{figure}[h]
\centering
\includegraphics[scale=0.27]{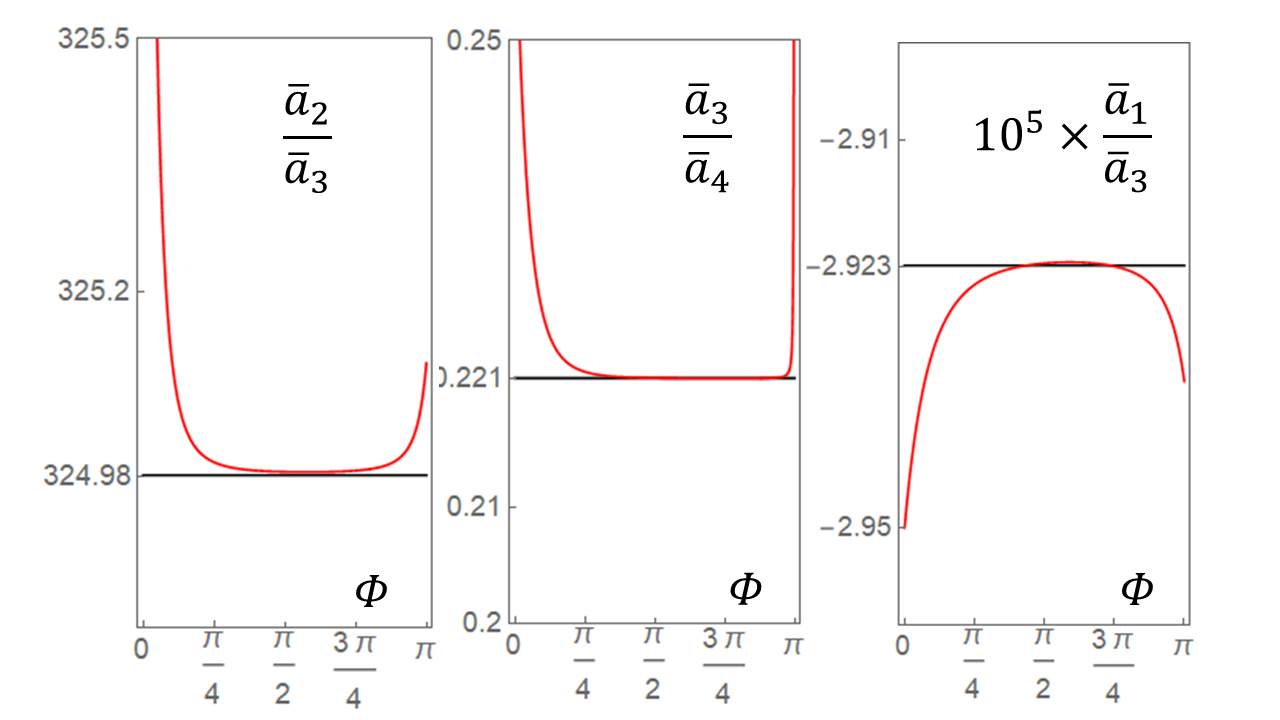} 
\caption{Matching ratios in Eq. (\ref{203}) with experimental parameters for hexagon  $\zeta=0.053$cm, $m=0.216, R_0=1.18$ cm, $p=3.44$,  see Table I, for the whole range . The red curves represent the left-hand sides of the terms in Eq. (\ref{203}) as functions of the poloidal angle $\phi$. The black lines represent the right-hand side terms of the same equation, obtained from experiments.  The theory-experiment match is acceptable for the large part of the range of the poloidal angle.}
\label{figextra} 
\end{figure}
We match cnoidal wave theoretical solution Eqs. (\ref{el},\ref{202}) with our experimental observation of spontaneously rotating hollow polygonal patterns, as described in section \ref{semodel1} and using Eqs. (\ref{201}-\ref{203}). The free parameter $\delta=\epsilon \sin \phi$ was chosen in each case (for each value $n=3, 4, 5, 6$ and for the most stable polygons in experiments) by minimizing the total difference between right-hand side terms in Eq. (\ref{203}) obtained in experiments and the theoretical expressions in the left-hand side depending on $\phi$, see Fig. \ref{figextra}. The matching procedure consists in numerically adjusting the parameters $\epsilon, a_0$ and $Re_{\Gamma}$ until the left- and right-hand side terms in Eq. (\ref{203}) are as close as possible along a range of $\phi$ as large as possible. The total difference was calculated as the integral over $[0,\pi/2]$ of the absolute value of the difference of the left and right terms, correspondingly. The errors occur at the ends of the interval for $\phi$ (at $0$ and $\pi/2$) because of the circular shape approximation made for the true shape of the vertical profile $\Sigma$. In Appendix III we present a proof of existence of a match between shape and the theoretical parameters.

The most relevant experimental results are presented in Figs. 
\ref{figs6543}-\ref{figs6543b} and in the Supplemental Materials \cite{supp}. Examples of numerical values for the parameters, for $R_0=1.2$ cm and   $h=0.3$ cm,   are shown in Table I. The corresponding match with the nonlinear cnoidal wave solution from Eqs. (\ref{el}-\ref{203}) is overlapped to the light green curve, in each figure. 
\begin{figure}[h]
	\centering
	\begin{tabular}{cc}
\includegraphics[scale=.22]{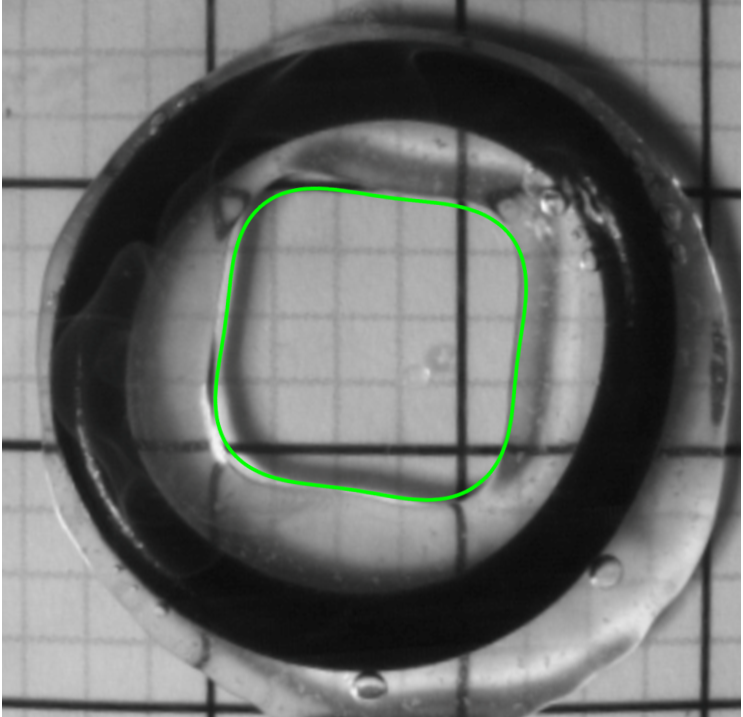} &
\includegraphics[scale=.22]{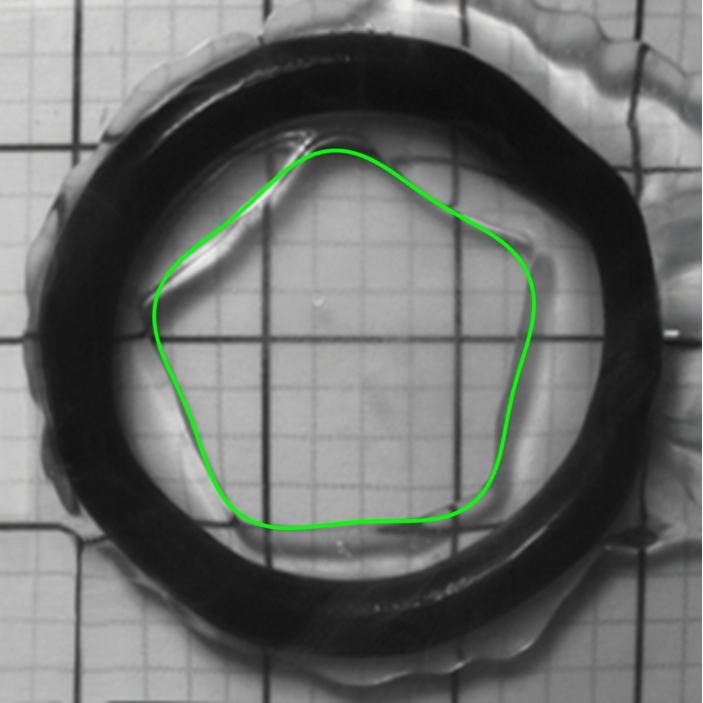} \\
	\end{tabular}
\caption{Experimental results $n=4,5$ matched with model Eq. (\ref{el}) and Eq. (\ref{eqfin2}), light green polygons, with seeding particles visible in the left frame. These patterns are also presented in Fig. \ref{figtheo} by the square and pentagon  with  $d/R=.21, \Omega\simeq 8.89$ s$^{-1}$ and $d/R=0.206, \Omega=6.76$ s$^{-1}$ , respectively. The external layer of liquid nitrogen is also observable. The edge of large squares $=1.27$ cm, so in the left frame $R=1.37$ cm, and in the right frame  $R=1.60$ cm.   See  Supplemental Materials \cite{supp}}
	\label{figs6543} 
\end{figure}
\begin{figure}[h]
	\centering
	\begin{tabular}{cc}
\includegraphics[width=4cm,height=3.8cm]{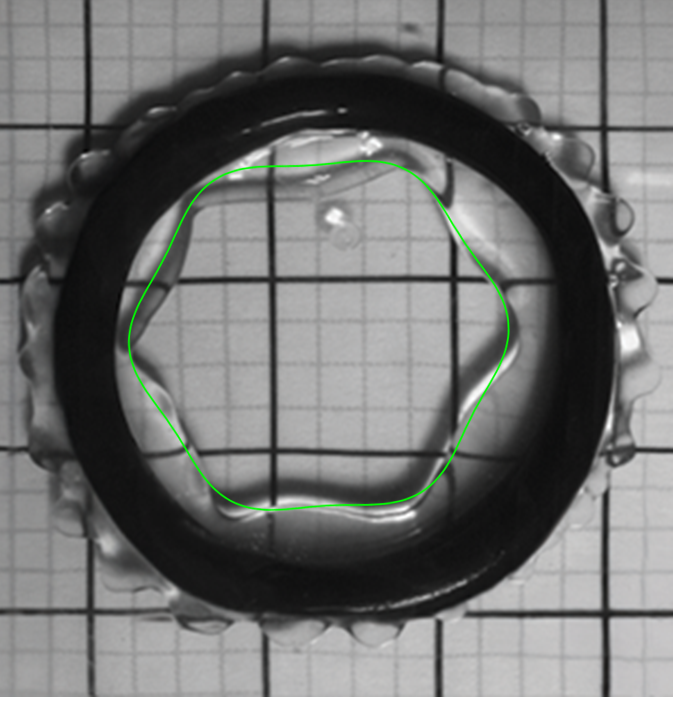} &
\includegraphics[width=4cm,height=3.8cm]{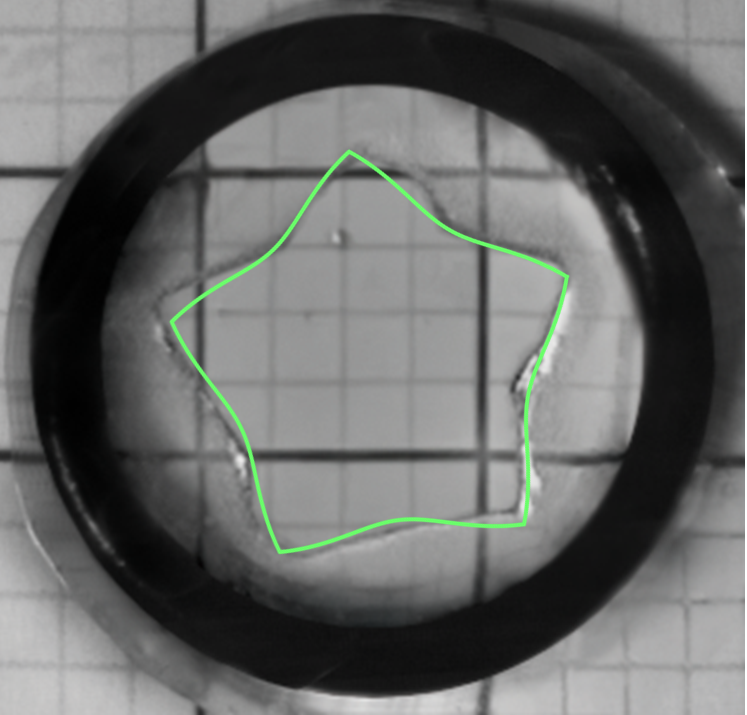} \\
	\end{tabular}
\caption{\textit{Left frame:} 
Hexagonal  pattern  $d/R=0.157, \Omega=7.01$ s$^{-1}$  matched with model Eq. (\ref{el}) and Eq. (\ref{eqfin2}) in light green, also schematically presented in Fig. \ref{figtheo}. 
\textit{Right frame:} Peaked polygonal pattern  $d/R=0.29, \Omega=11.8$ s$^{-1}$    matched with  solution in Eq. (\ref{peak}), light green curve. The edge of large squares $=1.27$ cm, so in both frames $R=1.40$ cm.   See  Supplemental Materials \cite{supp}.}
	\label{figs6543b} 
\end{figure}
\begin{center}
	\begin{table}[h]
		\centering
		\begin{tabular}{|c|c|c|c|}
			\hline\xrowht{10pt}
			Parameters& $n=4$  & $n=5$  & $n=6$ \\
			\hline\xrowht{10pt}
			 	$d \ [cm]$ (Exp.) &  $0.29\pm 0.01$ &  $0.33 \pm 0.01$ &  $0.22 \pm 0.01$   \\ 
			\hline\xrowht{10pt}
			$\Omega \ [s^{-1}]$ (Exp.)&  $8.89\pm 0.02$   &  $6.76\pm 0.02$ &  $7.01\pm 0.02$ \\
			\hline\xrowht{10pt}
			$\Omega \ [s^{-1}]$ (Th.)&  $9.16 $   &  $ 6.44$ &  $ 7.06$ \\
			\hline\xrowht{10pt}
			$Re_{\Gamma} $ (Exp.)&  $2170 \pm 260$   &  $1700 \pm 205$ &  $1650 \pm 201$ \\
			\hline\xrowht{10pt}
			$Re_{\Gamma} $ (Match)&  $2043$   &  $1775$ &  $1716$ \\
			\hline\xrowht{10pt}
			 	$a=\epsilon R_0 \ [cm]$ (Match)   &  $0.128$   &  $0.469$   &  $0.441$ \\	
			\hline\xrowht{10pt}
			 				$\zeta \ [cm]$ (Th.)   &  $0.118$ &  $0.082$ &  $0.053$ \\
			\hline\xrowht{10pt}
			$m$ (Th.)&  $0.402$   &  $0.414$ &  $0.216$ \\
			\hline\xrowht{10pt}
			$p$ (Th.) &  $2.37$   &  $3.04$ &  $3.44$ \\
			\hline
		\end{tabular}
		\caption{  Parameters $d, \Omega$ (Exp.), $Re_{\Gamma}$ (Exp.) are obtained from experiment,  $Re_{\Gamma}$ (Match), $a$  are the matching free parameters, and $\zeta, m, p$ are from the theoretical model for the specific polygon, respectively. The theoretical angular velocity $\Omega$ (Th.) is obtained from the nonlinear dispersion relation Eq. (\ref{204x}).   Experiment data is for liquid nitrogen rings with $\rho=807$ Kg/m$^3$, $\sigma=8.96 \cdot 10^{-3}$ N/m,   $h=0.3$ cm   and equilibrium radius   in the range $R_0=1.08-1.27$ cm. }
		\label{time}
	\end{table}
\end{center}
In Table I we present an example of the numerical results of the matching between the experiment and theoretical model for three polygons: square, pentagon and hexagon.   The quantities $R_0, h, d, \Omega$ (Exp.) in the first column are measured from each experiment. $Re_{\Gamma}$ (Exp.) is obtain with Eq. (\ref{vorty}) using   $h\sim 0.3$ cm,   $r_{\Gamma},R=R_0+d$ from experiments, and velocities $V_z\sim 0.13, \ V_r\sim 0.10$  m$\cdot$s$^{-1}$ from tracers displacements measurement, also experiment. $\Omega$ (Th.)   The free model parameters $a=\epsilon R_0, Re_{\Gamma}$ (Match) are obtained through the matching procedure described above. The cnoidal waves parameters $\zeta, m, p$ are obtained from the theoretical model, for each polygon shape, respectively.   The theoretical evaluation of the angular velocity of rotation of polygons $\Omega$ (Th.) is based on the nonlinear dispersion relation for cnoidal waves, presented in Eq. (\ref{204x}),  Appendix 2  . We notice a   good   correspondence between experiments and the predictions of the theoretical model. 

In some of the experiments, especially for the square patterns, we notice the occurrence of some type of shape instability. In some situations the polygon shapes become weakly unstable, and start to have slow and small amplitude curvature changes,  constraining the polygon edges to oscillate between concave and convex shapes. In other cases a smaller localized perturbation, like a secondary soliton solution overlapping the cnoidal wave, is generated at a vertex, and travels along the polygon edges with a higher that $\Omega$ angular velocity. When the edge of a polygon bends outwards the mean curvature pressure  tends to restore this effect and the polygonal shape begins to oscillate between a peaked (concave) and a round corners polygon,  around the equilibrium shape. When the $\Gamma$ contour perimeter tends to increase because of these deformations, the potential energy increases, and the  polygonal shape exhibits self-sustaining behavior.  

\section{Conclusions}
\label{secconcl}

We present a nonlinear model for the formation of hollow, rotating polygonal patterns in Leidenfrost rings of liquid nitrogen and compare the model’s predictions with previous experimental results \cite{rag}. Currently, no complete theoretical framework exists for predicting the spontaneous formation of sharp-cornered polygons in rotating Leidenfrost rings on a flat substrate. In this work, we model the phenomenon as cnoidal wave solutions to a fully nonlinear equation. Due to the complexity of the system, not all model parameters can be derived from first principles. Therefore, we introduce four free parameters, which are empirically determined from experimental data. This approach serves as the foundation for the current study, with the goal of providing a more comprehensive model that incorporates the relevant factors involved.

We identify the temperature regime and explore the potential for turbulent flow. The thermal properties of liquid nitrogen are examined, and the onset of Rayleigh-Bénard-Marangoni instability is discussed. The balance between buoyancy-driven and thermocapillary flows is also considered. Furthermore, we evaluate several dimensionless numbers relevant to the fluid dynamics of the system.  The paper includes elements of analysis of potential flow and vorticity, and the structure of the rolling vortex. A detailed calculation of the surface tension term is provided in Appendix I. 

To predict the formation of rotating polygonal patterns, we model the dynamics of the inner free surface of the Leidenfrost ring using the Euler equation for an incompressible fluid, within the Boussinesq shallow convection approximation. The fluid velocity is decomposed into potential and rotational components, with the corresponding Helmholtz-Hodge boundary conditions applied. The potential component of the velocity field is solved via the Laplace equation using a power series expansion in the radial coordinate. Due to the shallow nature of the liquid layer, we approximate the vorticity term as a gradient, allowing it to be included in an integrable form.

The resulting evolution equation for the free surface of the liquid ring, along with the boundary conditions, can be reduced to a single nonlinear differential equation for the contour curve $\gamma(\theta,t)$, which describes the shape of the inner free surface. To facilitate analysis, we solve this equation in the co-rotating frame, using a single variable $\xi=\theta-\Omega t$. Vorticity is incorporated into this equation through the vortex Reynolds number, treated as a free matching parameter. The surface tension term is derived using the mean curvature for the parametrization of the inner surface $\Sigma$, and expanded in a power series in small parameters related to the aspect ratio of the liquid ring. The final evolution equation contains twelve terms, with nonlinearities up to $\mathcal{O}(\gamma^3)$. An approximate analytical solution is obtained, and the resulting cnoidal wave solutions, expressed in terms of elliptic Jacobi functions, are analyzed,  with detailed calculations presented in Appendix II. We also conduct a qualitative analysis of this nonlinear differential equation.

We further validate the model by averaging the surface tension term and solving the resulting simplified evolution equation. We demonstrate that this simplification leads to a Korteweg-de Vries-type integrable equation, which yields the same cnoidal wave solutions, albeit in a different parametrization. If we implement in this simpler model  the contribution of the radial velocity we obtain a non-integrable nonlinear Eq. (\ref{fi3}). Surprisingly, we were able to obtain an exact analytic family of solutions for this  Eq. (\ref{fi3}) as the absolute value of a trigonometric function Eq. (\ref{peak}), which can match very well experiments resulting in  peaked polygonal patterns, like the example for $R_0/d=n=5$ presented in the right frame of Fig. \ref{figs6543b}. 

The cnoidal wave solutions Eq. (\ref{el}) are then matched to experimental data for polygonal patterns using the four free parameters, with a specialized procedure employed to account for variations in these parameters as a function of the poloidal angle $\phi$ of the vertical surface profile. See the 4 movies and 7 photographs as Supplemental Materials \cite{supp}. Further details on the matching comparison procedure outlined in Appendix III.

Our model presents several limitations that should be acknowledged. First, the flow calculations do not account for the viscous shear from the vapor escaping beneath the drop or its interaction with the internal dynamics, although this effect is expected to be weak for larger drops, as discussed in reference \cite{yimb}.
Additionally, we did not consider the precise height profile of the polygonal states (region $\Sigma$ in Fig. \ref{fig1}). Acquiring such data would require specialized high-speed video equipment.  The formation of rolling vortices was incorporated only through an average contribution represented by the vortex Reynolds number $Re_{\Gamma}$. Moreover, our experiments did not measure the temperature distribution across the Leidenfrost ring at various points, limiting the accuracy of our evaluation of the Marangoni effect. Changes in the contact angle due to rapid rotation were also not included in our model.  In addition, in a more advanced model describing longer duration of rotations and oscillation,  weak viscosity and  thin-film lubrication terms might be included as small damping terms.

The primary focus of our current model is to elucidate how nonlinear terms in the evolution equation contribute to the formation of the observed rotating polygons. However, to obtain a more comprehensive understanding of both the internal dynamics and the spontaneous formation of rotating polygonal patterns, it would be beneficial to extend our model by incorporating thermo-convective effects in a generalized three-dimensional framework. Furthermore, it would be valuable to investigate the stability of different polygonal shapes and the transitions between polygons with varying numbers of edges as liquid nitrogen evaporates and the equilibrium radius $R_0$ decreases.  Such extensions would facilitate exploration of the conditions under which polygonal shapes emerge and how they transition between different forms. As recommended by the authors of similar studies \cite{2012bohr,bohr2011,bohr2013,bohr2019,jansson2006,28,amauche2013,hamid,vatistas90,abderrah2017}, it is anticipated that such research will become available in the near future. One of the key objectives of this paper is to encourage further investigation in this direction.

\appendix
\numberwithin{equation}{section}
\renewcommand{\theequation}{A1.\arabic{equation}}
\section*{Appendix 1}
\label{secapp1}
\setcounter{equation}{0}

\begin{figure}[H]
\centering
\includegraphics[scale=0.31]{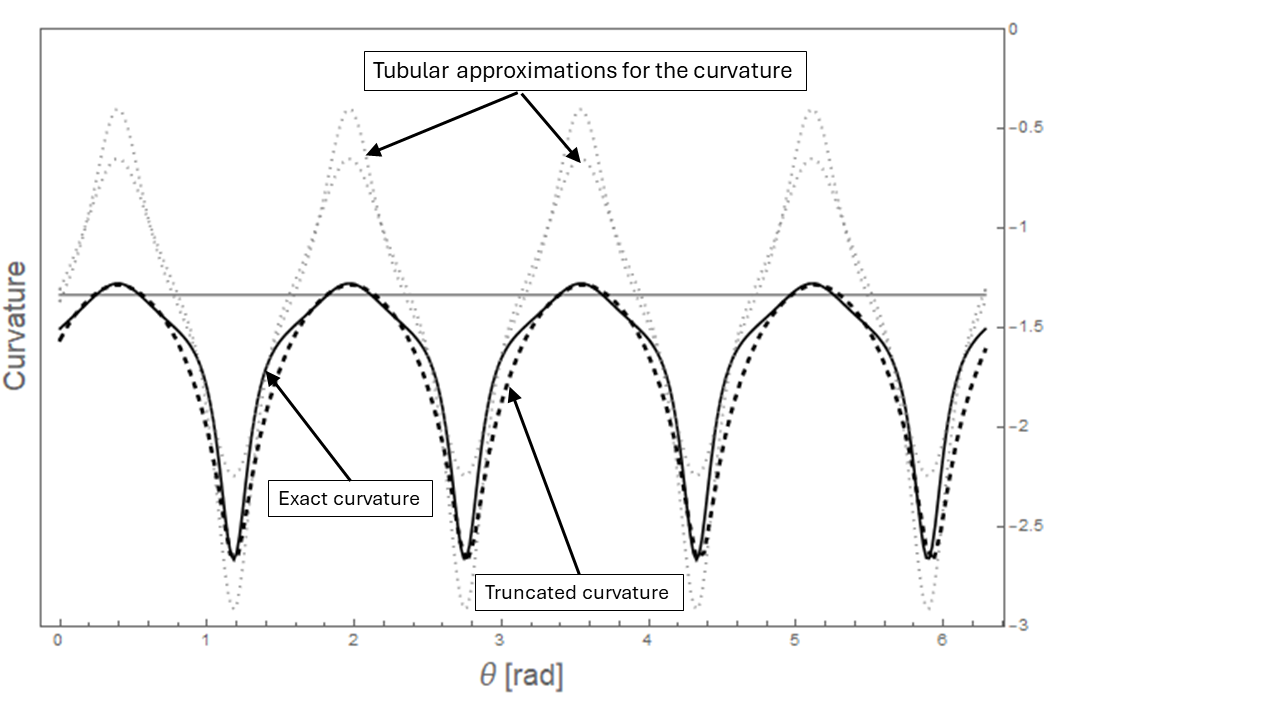} 	
\caption{Plot of the mean curvature $H_{\Sigma}(\theta, \varphi)$ for a square shaped contour $\Gamma$,  for $\varphi=\pi/4$  poloidal angle, $R_0=2$cm,   $a=0.3$cm.   The solid line represents the exact expression of the mean curvature from Eq. (\ref{A1e3}) and the dashed curve represent the  truncated expression  Eqs. (\ref{eqp}-\ref{A1e4}). For comparison, we plot mean curvature expression given by the tubular regular surface Eq. (\ref{A1e2}), dotted gray curves. The horizontal line represents a constant mean curvature $H_0=(1/R_0-1/a)/2$ in a case of a static equilibrium configuration with $\gamma=0$, that is a half-torus.}
\label{figcurv}
\end{figure}

In order to evaluate the surface tension pressure at the inner liquid boundary $\Sigma$  we use the Young-Laplace Eq. (\ref{e11}), so we need to calculate the mean curvature $H_{\Sigma}$ of this surface. As described above, we model $\Sigma$ as a normal ringed  surface of poloidal radius $a$, generated by a circular arc following in the normal plane the $\Gamma$ contour Eq. (\ref{ega}). 
This surface is parameterized in Eq. (\ref{A1e1}). If the curvature $\kappa_{\Gamma}$ of the $\Gamma$ contour obeys the condition $\kappa_{\Gamma}\le 1/a$ the surface $\Sigma$ is regular, and its mean curvature has a simple form \cite{tubu}
\begin{equation}\label{A1e2}
H_{\Sigma, \hbox{reg}}=\frac{1}{2}\biggl( -\frac{1}{a}+\frac{\kappa_{\Gamma} \cos \varphi}{1- a \kappa_{\Gamma} \cos \varphi} \biggr).
\end{equation}
However, since the patterns of interest generated by $\Gamma$ are polygons with sharp corners, the surface $\Sigma$ is not regular in some neighborhood of the vertices, and the mean curvature in Eq. (\ref{A1e2}) acquires singularities at the vertices. For example, such irregular regions with  $\kappa_{\Gamma}> 1/a$ are shown in  Fig. \ref{fig2} by the red colored areas, around the vertices. Because of these situations we need to calculate the mean curvature from its definition, using the first and second fundamental forms of $\Sigma$ \cite{x3,book,tubu}
\begin{equation}\label{A1e3}
H_{\Sigma}=\frac{eG-2fF+gE}{2(EG-F^2)},
\end{equation}
where $E, F, G$ and $e,f,g$ are the coefficients of the first and second fundamental forms, respectively
$$
E=\vec{r}_{\Sigma,\theta}\cdot \vec{r}_{\Sigma,\theta}, \ F=\vec{r}_{\Sigma,\theta}\cdot \vec{r}_{\Sigma,\varphi}, \ G=\vec{r}_{\Sigma,\varphi}\cdot \vec{r}_{\Sigma,\varphi},    
$$
$$
e=\vec{r}_{\Sigma,\theta \theta}\cdot \vec{N}_{\Sigma}, \  f=\vec{r}_{\Sigma,\theta \varphi}\cdot \vec{N}_{\Sigma}, \ g=\vec{r}_{\Sigma,\varphi \varphi}\cdot \vec{N}_{\Sigma},
$$
where subscripts represent differentiation and $\vec{N}_{\Sigma}$ is the unit normal to the surface \cite{x3}. The calculations are tedious and long, and we present the Mathematica\textsuperscript{\textregistered} code for calculating Eq. (\ref{A1e3}) in Supplemental Materials \cite{supp} the file Curvature.nb. If we expand the exact form of Eq. (\ref{A1e3}) for $H_{\Sigma}$ in power series with respect to $a, \gamma$ and retain terms up to order three, we obtain
\begin{equation}\label{eqp}
\begin{split}
H_{\Sigma} & \simeq A_0+A_1 \gamma+A_2 \gamma^2+A_3 \gamma^3+B_1 \gamma_{\theta}^{2}
+B_2 \gamma_{\theta \theta} \\
& +C_1 \gamma \gamma_{\theta \theta}+C_2 \gamma \gamma_{\theta}^{2},
\end{split}
\end{equation}
where the coefficients of various terms in $\gamma$ and its derivatives are functions of the angle $\varphi$. We define two smallness parameters 
\begin{equation}\label{A1e3p}
\epsilon=\frac{a}{R_0}, \ \  \delta=\frac{a \sin \varphi}{R_0},
\end{equation}
and with these notations, the coefficients in Eq. (\ref{eqp}) have the following expressions
\begin{widetext}
\begin{eqnarray}\notag
A_0 & = \frac{1}{2 a} [-1+\delta+\delta^2-\delta \epsilon-2\delta^2 \epsilon+\delta \epsilon^2+3 \delta^2 \epsilon^2-\delta \epsilon^3 + \mathcal{O}_4(\epsilon,\delta)
] \\
\notag
A_1 & = \frac{1}{2 a} [-\delta-2\delta^2+2\delta \epsilon+6\delta^2 \epsilon-3\delta \epsilon^2-12 \delta^2 \epsilon^2+4\delta \epsilon^3 + \mathcal{O}_4(\epsilon,\delta)
] \\
\notag
A_2 & =\frac{1}{2 a} [\delta+3\delta^2-3\delta \epsilon-12\delta^2 \epsilon+6\delta \epsilon^2+30 \delta^2 \epsilon^2-10 \delta \epsilon^3 + \mathcal{O}_4(\epsilon,\delta)
 ] \\
\label{A1e4}
A_3 & =\frac{1}{2 a} [-\delta-4\delta^2+4\delta \epsilon+20\delta^2 \epsilon-10\delta \epsilon^2-60 \delta^2 \epsilon^2+20\delta \epsilon^3 + \mathcal{O}_4(\epsilon,\delta)
 ] \\
\notag
B_1 & =\frac{1}{2 a} \biggl[\frac{\delta}{2}+\delta^2-\delta \epsilon-3\delta^2 \epsilon+\frac{3\delta \epsilon^2}{2}+6 \delta^2 \epsilon^2-2\delta \epsilon^3 + \mathcal{O}_4(\epsilon,\delta)
 \biggr], \  B_2=A_1 \\
\notag
C_1 & =\frac{1}{2 a} [2\delta+6\delta^2-6\delta \epsilon-24\delta^2 \epsilon+12\delta \epsilon^2+60 \delta^2 \epsilon^2-20\delta \epsilon^3 + \mathcal{O}_4(\epsilon,\delta)
 ] \\
\notag
C_2 & =\frac{1}{2 a} \biggl[-\frac{3\delta}{2}-4\delta^2+4\delta \epsilon+15\delta^2 \epsilon-\frac{15\delta \epsilon^2}{2}-36 \delta^2 \epsilon^2+12\delta \epsilon^3 + \mathcal{O}_4(\epsilon,\delta)
 \biggr]
\end{eqnarray}
\end{widetext}
In order to verify the validity of this truncated expression for the mean curvature of $\Sigma$, we compare the  results of Eqs. (\ref{eqp}-\ref{A1e4})  with the exact form for the mean curvature from Eq. (\ref{A1e3}), and with the the tubular surface mean curvature approximation for regular surfaces Eq. (\ref{A1e2}). The comparison is presented in Fig. \ref{figcurv} for a particular example of a square polygon and $\varphi=\pi/4$. Because we choose a small capillary radius $a=3$mm in this example, the mean curvature is negative everywhere around the contour. Minimum negative values for $H_{\Sigma}$  are obtained at the corners of the square. A similar example is presented for a hexagon in Fig. \ref{fig2} where the negative curvature is colored in red.
The approximation given by the  tubular regular surface mean curvature expression Eq. (\ref{A1e2}) does not work well for our geometry; the gray dotted curves in Fig. \ref{figcurv} provide largely different results than the correct values around the vertices. We performed the same comparison for different values of  the poloidal angle $\varphi $ in the $[0,\pi]$ range, also for pentagon, hexagon and octagon $\gamma$ shapes. For all these cases we obtain no more than $5\%$ relative difference between truncated and full expression for the mean curvature.

\appendix
\numberwithin{equation}{section}
\renewcommand{\theequation}{A2.\arabic{equation}}
\section*{Appendix 2}
\label{secapp2}
\setcounter{equation}{0}

We write Eq. (\ref{eq5}) in a compact form
$$
\mathcal{A}_1\gamma'+\mathcal{A}_2\gamma\gamma'+\mathcal{A}_3\gamma^2\gamma'+\mathcal{A}_4\gamma'\gamma''+\mathcal{A}_5\gamma'''
$$
\begin{equation}\label{e00}
+\mathcal{A}_6(\gamma \gamma'')'+\mathcal{A}_7(\gamma \gamma'^{2})'=0,
\end{equation}
where the coefficients $\mathcal{A}_i$ are functions of  the physical parameters of the system: $We$ (or the angular velocity $\Omega$), $R_0, d, h, \rho, \sigma$, and also on three free parameters $\epsilon Re_{\Gamma}$ and $\delta$ (from Eq. (\ref{A1e3p})). Eq. (\ref{e00}), which models the evolution of the contour $\gamma(\xi)$ in the co-rotating frame, can be mapped into a perturbed KdV equation \cite{lamb}, by integrating it with respect to $\xi$, and then multiplying the result with a factor of the form $\gamma'e^{-\alpha \gamma}$. This transformation leads to an integrable form
\begin{equation}\label{eq6}
[(a_0+a_1 \gamma +a_2 \gamma^{2}+a_3 \gamma^3+a_4 \gamma'^{2}+a_5 \gamma \gamma'^{2}) e^{-\alpha \gamma}]'=0,
\end{equation}
where the six coefficients $a_0,\dots , a_5$, the auxiliary constant $\alpha$ 
are obtained as functions of the coefficients $\mathcal{A}_1, \dots,\mathcal{A}_7$, through identification of the independent  terms 
between Eq. (\ref{e00}) and (\ref{eq6}).  From integration of Eq. (\ref{eq6}) we obtain the traveling-wave reduction of the KdV equation, perturbed with a quasi-linear Hamiltonian nonlinearity \cite{lamb,perturb} in the right-hand side
\begin{equation}\label{e13}
a_0+a_1 \gamma +a_2 \gamma^{2}+a_3 \gamma^3+a_4 \gamma'^{2}=-a_5 \gamma \gamma'^{2}.
\end{equation}
Differentiating Eq. (\ref{e13}), dividing by $\gamma'$, and differentiating again results in  a new expression
$$
2 a_2 \gamma'+6 a_3 \gamma \gamma'+2 a_4 \gamma^{'''}=-a_5 (\gamma^{'2}+2 \gamma \gamma{''})^{'},
$$
which can be written in Hamiltonian formalism
\begin{equation}\label{e14}
2 a_2 \gamma^{'} =-\frac{2 a_2}{\Omega}\gamma_t= -\frac{\partial}{\partial_{\xi}} \nabla H(\gamma) 
	=\frac{\partial}{\partial_{\xi}} \biggl( \frac{\partial \mathcal{H}}{\partial \gamma}-\frac{\partial }{\partial \xi}\frac{\partial \mathcal{H}}{\partial \gamma_{\xi}} \biggr),
\end{equation}
with
$$
H=\int_{0}^{2 \pi} \mathcal{H} d\xi=\int_{0}^{2 \pi} ( a_4 \gamma_{\xi}^2-a_3 \gamma^3 +a_5 \gamma \gamma_{\xi}^2 ) d\xi. 
$$
For this type of quasi-linear Hamiltonian equation it was shown \cite{kam} that KdV soliton solutions are only affected by small-amplitude, quasi-periodic perturbations.

In the following we show how the order of the perturbation in Eq. (\ref{e13}) can be further reduced to $\mathcal{O}(\gamma^4)$  leading  to the $\mathcal{O}(\gamma^3)$ non-perturbed KdV equation 
\begin{equation}\label{e15}
\bar{a}_0+\bar{a}_1 \gamma +\bar{a}_2 \gamma^{2}+\bar{a}_3 \gamma^3+\bar{a}_4 \gamma'^{2}=0.
\end{equation}
The proof is straightforward: we assume $\gamma$ is a solution of Eq. (\ref{e15}) and then implement this constraint in  Eq. (\ref{eq6}). By identification we obtain
\begin{equation}\label{lin}
\begin{split}
& \bar{a}_0=a_0, \ \bar{a}_1=a_1-\frac{a_5 a_0}{a_4}, \  \bar{a}_2=a_2-\frac{a_5}{a_4}\biggl( a_1-\frac{a_0 a_5}{a_4} \biggr) \\ 
& \bar{a}_3=a_3-\frac{a_5}{a_4}\biggl( a_2-\frac{a_1 a_5}{a_4}+\frac{a_0 a_5^2}{a_4^2} \biggr), \  \bar{a}_4=a_4
\end{split}
\end{equation} 
The dependence of these $\bar{a}_i$ coefficients on the original physical parameters in Eq. (\ref{eq5}), through the expressions $\mathcal{A}_i$ in Eq. (\ref{e00}), is represented in terms of rational functions, yet  too long expressions to be included here. We present these expressions in the Supplemental Material \cite{supp} entitled "Coefficients.nb" where a Mathematica\textsuperscript{\textregistered} code calculates them.

The KdV Eq. (\ref{e15}) is a faithful representation of Eq. (\ref{eq5}) up to order $\mathcal{O}(\epsilon^4,\delta^4)$ and it has an exact elliptic function solution dubbed the cnoidal wave \cite{lamb} in the form
\begin{equation}\label{e20}
\gamma(\xi)=s_2+(s_3-s_2){\rm cn^2}[p(\xi-s_3)|m],
\end{equation}
where  $s_1\le s_2 \le s_3$ are the roots of the algebraic equation 
\begin{equation}\label{eqpolyn}
-\frac{\bar{a}_3}{\bar{a}_4} \biggl( s^3+\frac{\bar{a}_2}{\bar{a}_3}s^2+\frac{\bar{a}_1}{\bar{a}_3}s+\frac{\bar{a}_0}{\bar{a}_3}\biggr)=0,
\end{equation}
and the scaling factors are
\begin{equation}\label{eqparame}
p=\frac{1}{2}\sqrt{\frac{\bar{a}_3(s_3-s_1)}{\bar{a}_4}}, \ \  m=\frac{s_2-s_3}{s_1-s_3}\in [0,1].
\end{equation}
Incorporating the relations Eqs. (\ref{201}, \ref{202}) together with  Eqs. (\ref{eqpolyn},\ref{eqparame}) for the scaling factors we obtain
\begin{equation}\label{203}
\frac{\bar{a}_2}{\bar{a}_3}=\frac{\zeta (2-m)}{m R_0}, \  \frac{\bar{a}_3}{\bar{a}_4}=\frac{4 p^2 m R_0}{\zeta}, \ \frac{\bar{a}_1}{\bar{a}_3}=-\frac{\zeta^2}{R_{0}^{2}}. 
\end{equation}
  It results that for a choice of $R_0$ and $\Omega$ we can fit a certain rotating polygonal shape for the $\Gamma$ contour,  which is uniquely determined by $\zeta, m, p$ and $\Omega$ according to Eq. (\ref{202}), with the values of the rations $\bar{a}_2 / \bar{a}_3, \bar{a}_3 / \bar{a}_4$ present in Eq. (\ref{eqpolyn}). Indeed, the above coefficients $\bar{a}_1, \dots, \bar{a}_4$ are completely determined by the coefficients $\mathcal{A}_1, \dots, \mathcal{A}_7$ from Eq. (\ref{e00}). Through Eq. (\ref{eq5}) the coefficients $\mathcal{A}_i$ depend on measurable physical parameters of the system and on the four free parameters of the model  $Re_{\Gamma}, a_0$ and $\epsilon, \delta$.   In conclusion, the three equations Eqs. (\ref{203}) make possible the one-to-one match between shape and the free parameters $\{ \zeta,p,m \} \leftrightarrow \{ Re_{\Gamma},  a_0, \epsilon, \delta \}$. The existence of solutions for this match is proved in Appendix III. The last free term  inside the parenthesis in Eq. (\ref{eqpolyn}) results from the geometry parameters 
\begin{equation}\label{222}
\frac{\bar{a}_0}{\bar{a}_3}=-\frac{\zeta^3 (2-m)}{m R_{0}^{2}},
\end{equation}
because $\bar{a}_0$ is a free constant  resulting from the integration of Eq. (\ref{eq6}). 

The solution Eq. (\ref{e20}) is periodic of period $4 K(m)/p$ with $K(\cdot)$ being the elliptic integral of the first kind, and it oscillates between $s_2$ and $s_3$. In the linear limit $cn^2(p \xi|0)=\cos^2 p \xi$ and in the limit  $cn^2(p \xi|1)=\hbox{sech}^2 p \xi$ the cnoidal wave becomes a KdV soliton \cite{lamb}.

Within the model we can estimate  the angular velocity $\Omega$ from the nonlinear dispersion relation for nonlinear waves  \cite{whit}, \cite{kevr} or section 19.5 in \cite{book}.   We map the differential equation Eq. (\ref{eq5}) into the algebraic equation for the nonlinear dispersion relation $\Omega(\lambda)$ by the substitutions $\{ \gamma, \gamma', \gamma'', \dots \} \rightarrow \{ \zeta/R_0, \zeta/(R_0 \lambda), \zeta/(R_0 \lambda^2), \dots \}$ and obtain
\begin{equation}\label{204x}
\Omega\simeq \sqrt{\frac{\sigma}{\rho h^3} \cdot  \frac{R^4 a \delta d^2(1-\zeta+\lambda^2-2 \zeta \lambda^2)}{[3 \zeta d^2+R(\zeta R-d)\lambda^2+4 \zeta  h d Re_{\Gamma}] h^5}}
\end{equation}
We plug the corresponding values for the quantities $a, \delta, d, \zeta$ for each $n-$polygon with the angular wavelength $\lambda=2 \pi R_0 / n$, and we present the results in the third row 	$\Omega \ [s^{-1}]$ (Th.)	of the Table \ref{time}.

\appendix
\numberwithin{equation}{section}
\renewcommand{\theequation}{A3.\arabic{equation}}
\section*{Appendix 3}
\label{secapp3}
\setcounter{equation}{0}
\begin{figure}[h!tbp]
	\centering
	\includegraphics[scale=0.4]{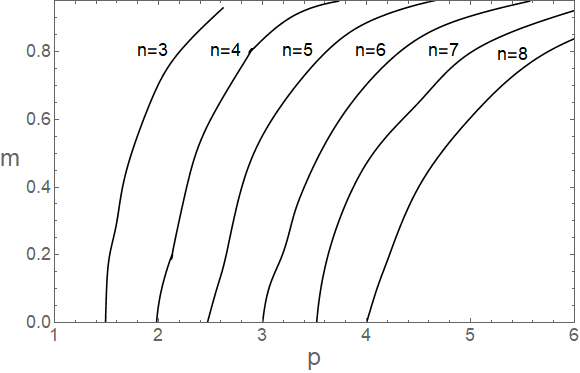} 	
	\caption{Plot of the values of the parameter $m$ function of the parameter $p$ from the cnoidal waves solution in Eq. (\ref{el}), or similarly Eqs. (\ref{202},\ref{e20}), which realize a polygonal contour. That is the curves $m=m(p,n=3),\dots, m=m(p,n=8)$.}	
	\label{fig1001}
\end{figure}

The solution Eq. (\ref{202}) representing polygonal shapes is periodic around its base circle of radius $R_0$ for specific values of the parameters $p,m$, while the roots $s_2,s_3$ of the polynomial equation Eq. (\ref{eqpolyn}) contribute only to the radial amplitude $\zeta$ of the contour deformations. Performing a numerical analysis on the admissible values for these $p,m$ parameters and for the Jacobi elliptic cosine polygonal solution, we obtained the following result presented in Fig. \ref{fig1001}: For any type of polygonal contour $\Gamma$ with $n=3,\dots,8$ edges represented by this solution in Eq. (\ref{el}), or similarly Eqs. (\ref{202},\ref{e20}), and for any value for $m\in [0,1]$ there is a unique value for $p$ which can represent that polygon, given by the intersection of the horizontal line $m=$constant with any of the parametric elliptic curves $m=m(p,n)$ in Fig. \ref{fig1001}. This result offers an easier procedure to match the experimentally observed rotating polygons with the values $m,p$ and further to fit these shapes with the physical parameters of the model that predict such cnoidal waves solutions.

\begin{figure}[h!tbp]
	\centering
	\includegraphics[scale=0.265]{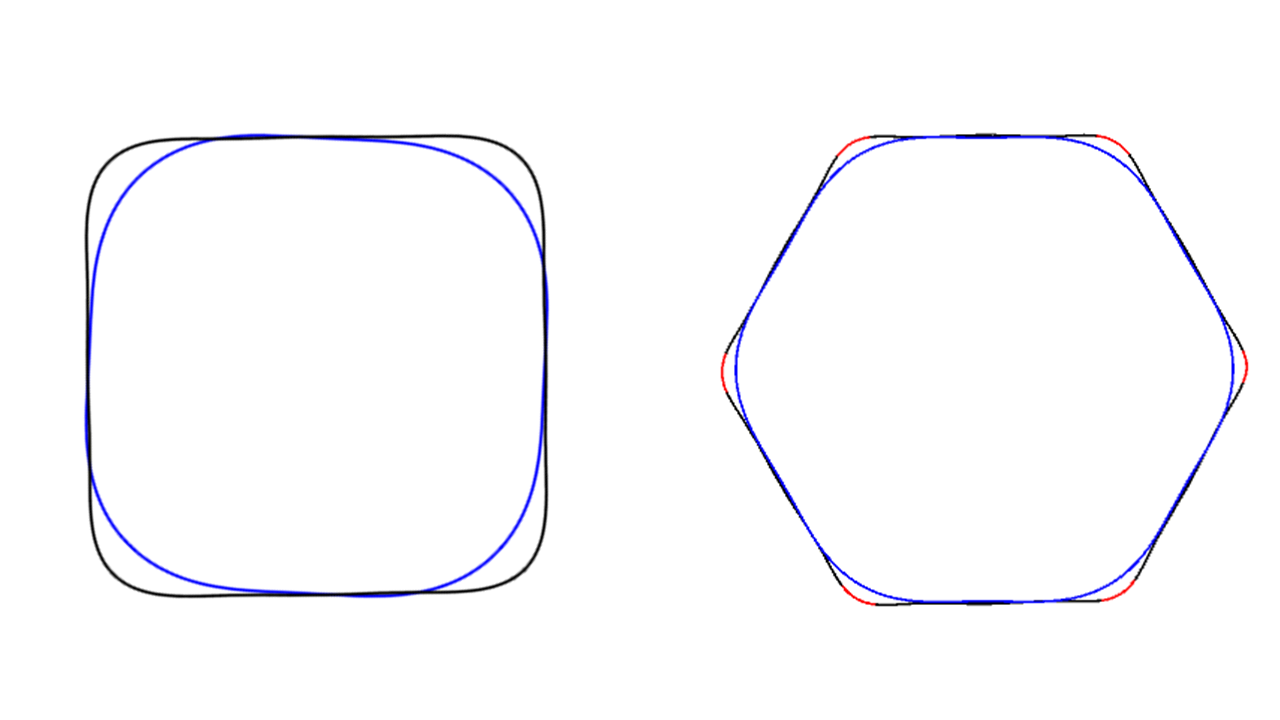} 	
	\caption{Comparison between linear contour waves (blue) and the corresponding cnoidal wave solutions (black, Eq. \ref{el}), for the same polygon modeling. The red vertices of the hexagonal shape represent regions where the curvature of cnoidal wave exceeds $1/a$, therefore canceling of mean curvature $H_{\Sigma}$ of the liquid surface.}
	\label{fig100}
\end{figure}

Another observation refers to the specifics of the nonlinear cnoidal solutions as compared to the linear solutions of this model. The solution Eq. (\ref{e20}) approaches in the linear limit $m\rightarrow 0$ the trigonometric solution $cn^2(p \xi|0)=\cos^2 p \xi$  \cite{lamb} which presents similarities with the polygonal contours, Fig. \ref{fig100}.
Nevertheless, the linearized solutions offer polygons with very rounded corners, represented by the blue curves in Fig. \ref{fig100} for certain values for $p$. In the large majority of the experiments we performed the rotating polygonal shapes have sharper corners, closer to the contours drawn by black curves. This comparison offers an extra argument in favor of the validation of the solutions obtained through the nonlinear model.

\end{document}